\documentclass[12pt]{article}
\usepackage[a4paper, top=16mm, text={170mm, 248mm}, includehead, includefoot, hmarginratio=1:1, heightrounded]{geometry}
\usepackage{amsmath,amssymb,mathrsfs,amsthm,tikz,shuffle,paralist}
\usepackage{dsfont}
\usepackage{color}
\definecolor{darkred}{RGB}{173,34,48}
\usetikzlibrary{snakes}
\usetikzlibrary{calc}
\usetikzlibrary{decorations}

\usepackage[all]{xypic}
\usepackage{jheppub}
\usepackage{hyperref}
\usepackage{subfigure}
\usepackage{caption}
\def\be{\begin{equation}}
\def\ee{\end{equation}}
\def\ba{\begin{eqnarray}}
\def\ea{\end{eqnarray}}

\def\Li{\textrm{Li}}
\def\l{\langle}
\def\r{\rangle}

\def\Roneone{R$^{1,1}$~}

\title{Bootstrapping octagons in reduced kinematics from $A_2$ cluster algebras}

\date{\today}
\author[a,b,c,d,e]{Song He,}
\author[a,e]{Zhenjie Li,} 
\author[a,e,f]{Yichao Tang,}%
\author[a,e]{Qinglin Yang}%
\affiliation[a]{CAS Key Laboratory of Theoretical Physics, Institute of Theoretical Physics, Chinese Academy of Sciences, Beijing 100190, China}
\affiliation[b]{Peng Huanwu Center for Fundamental Theory, Hefei, Anhui 230026, P. R. China}
\affiliation[c]{
School of Fundamental Physics and Mathematical Sciences, Hangzhou Institute for Advanced Study, UCAS, Hangzhou 310024, China}
\affiliation[d]{ICTP-AP
International Centre for Theoretical Physics Asia-Pacific, Beijing/Hangzhou, China}
\affiliation[e]{School of Physical Sciences, University of Chinese Academy of Sciences, No.19A Yuquan Road, Beijing 100049, China}
\affiliation[f]{School of Physics, Peking University, Beijing 100871, China}
\emailAdd{songhe@itp.ac.cn}
\emailAdd{lizhenjie@itp.ac.cn}
\emailAdd{tangyichao@itp.ac.cn}
\emailAdd{yangqinglin@itp.ac.cn}

\abstract{Multi-loop scattering amplitudes/null polygonal Wilson loops in ${\cal N}=4$ super-Yang-Mills are known to simplify significantly in reduced kinematics, where external legs/edges lie in an $1+1$ dimensional subspace of Minkowski spacetime (or boundary of the $\rm AdS_3$ subspace). Since the edges of a $2n$-gon with even and odd labels go along two different null directions, the kinematics is reduced to two copies of $G(2,n)/T \sim A_{n{-}3}$. In the simplest octagon  case, we conjecture that all loop amplitudes and Feynman integrals are given in terms of two overlapping $A_2$ functions (a special case of two-dimensional harmonic polylogarithms): in addition to the letters $v, 1+v, w, 1+w$ of $A_1 \times A_1$, there are two letters $v-w, 1- v w$ mixing the two sectors but they never appear together in the same term; these are the reduced version of four-mass-box algebraic letters. Evidence supporting our conjecture includes all known octagon amplitudes as well as new computations of multi-loop integrals in reduced kinematics. By leveraging this alphabet and conditions on first and last entries, we initiate a bootstrap program in reduced kinematics: within the remarkably simple space of overlapping $A_2$ functions, we easily obtain octagon amplitudes up to two-loop NMHV and three-loop MHV. We also briefly comment on the generalization to $2n$-gons in terms of $A_2$ functions and beyond.
}
\begin{document}

\maketitle

\section{Introduction and review}

Recent years have witnessed enormous progress in unravelling rich mathematical structures of scattering amplitudes in Quantum Field Theory (QFT). A remarkable example is ${\cal N}=4$ super-Yang-Mills (SYM), especially the theory in the planar limit: the all-loop integrand has been determined purely with on-shell data~\cite{ArkaniHamed:2010kv} and reformulated geometrically in terms of the positive Grassmannian~\cite{Arkani-Hamed:2016byb} and the amplituhedron~\cite{Arkani-Hamed:2013jha}; the (post-integration) amplitudes have been determined to very high loop orders for six and seven points (see~\cite{Dixon:2011pw,Dixon:2014xca,Dixon:2014iba,Drummond:2014ffa,Dixon:2015iva,Caron-Huot:2016owq,Dixon:2016nkn,Drummond:2018caf, Caron-Huot:2019vjl, Caron-Huot:2019bsq, Dixon:2020cnr,Chicherin:2017dob} and the review~\cite{Caron-Huot:2020bkp}). The starting point of hexagon and heptagon bootstrap is the observation~\cite{Golden:2013xva} that the {\it symbol alphabet}~\cite{Goncharov:2010jf, Duhr:2011zq} of six- and seven-point amplitudes are dictated by cluster algebras (see ~\cite{fomin2002cluster,fomin2003cluster}) naturally associated with $G(4,6)/T$ and $G(4,7)/T$, i.e., $A_3$ and $E_6$, respectively~\cite{speyer2005tropical}. Starting at eight points, $G(4,n)/T$ cluster algebras are infinite and the (finite) symbol alphabets involve algebraic letters that go beyond the usual cluster coordinates. Recently, the two-loop NMHV amplitudes\footnote{As this paper is being prepared, the 3-loop MHV octagon amplitude has also been computed using the $\bar Q$ equation in~\cite{newpaper}.} have been computed for $n=8$~\cite{Zhang:2019vnm} and higher~\cite{He:2020vob} using the method of $\bar Q$ equations~\cite{CaronHuot:2011kk}, and the alphabet has been explained using tropical positive Grassmannians~\cite{Drummond:2019cxm, Henke:2019hve, Arkani-Hamed:2019rds} (see also~\cite{Herderschee:2021dez}) as well as Yangian invariants and plabic graphs~\cite{Mago:2020kmp, He:2020uhb,Mago:2020nuv}. There has also been new progress on the cluster algebra structures for individual Feynman integrals in ${\cal N}=4$ SYM~\cite{Caron-Huot:2018dsv,He:2021esx} and in a broader context~\cite{Chicherin:2020umh}.

Despite the hidden simplicity and rich structures of ${\cal N}=4$ SYM, it quickly becomes too difficult to compute or understand scattering amplitudes and Feynman integrals in an analytic form, as the number of legs and loops increase. For example, beyond seven points, the functions for two-loop NMHV amplitudes and double-pentagon integrals, whose symbols involve algebraic letters~\cite{He:2020vob,He:2020lcu}, have not been determined so far (let alone those at higher loops). Moreover, the origin of these algebraic letters, the finer structures of the symbol such as cluster adjacency~\cite{Drummond:2017ssj,Drummond:2018caf}, and the patterns of algebraic words~\cite{Zhang:2019vnm,He:2020vob} have not been understood either. Given these limitations, it has become increasingly difficult to understand the cluster algebras and algebraic letters underlying these multi-leg amplitudes and integrals, or to compute/bootstrap them using such information. 

\def\Roneone{R$^{1,1}$~}

This is one of the motivations for considering restricted kinematical configurations where scattering amplitudes and Feynman integrals simplify. In this paper, we consider the so-called \Roneone kinematics, where external momenta or Wilson-loop edges lie in an $1+1$ dimensional subspace of the Minkowski spacetime. It is well known that MHV amplitudes/bosonic Wilson loops simplify a lot in \Roneone kinematics. In fact, \Roneone Wilson loops were first computed at strong coupling via AdS/CFT correspondence~\cite{Alday:2009ga,Alday:2009yn} since the minimal surfaces simplify greatly in the $\rm AdS_3$ subspace; later they were computed at weak coupling as well~\cite{DelDuca:2010zp,Heslop:2010kq}. In~\cite{Caron-Huot:2013vda}, this was extended to non-MHV amplitudes (supersymmetrically) reduced to \Roneone kinematics: by combining the all-loop $\bar Q$ equations~\cite{CaronHuot:2011kk} and a collinear-soft uplifting formalism~\cite{Goddard:2012cx}, compact analytic formulas with $k+\ell=3$ (i.e., one-loop N$^2$MHV, two-loop NMHV and three-loop MHV) were obtained for the first time for octagons and essentially extends to all multiplicities! We believe that these results have only revealed a tiny portion of the richness of scattering amplitudes in \Roneone kinematics: not only have we obtained very compact formulas with huge simplifications at higher loops and deeper into the non-MHV sectors, but we also see non-trivial structures. Therefore, although simpler, \Roneone kinematics still exhibits considerable complexity, making it an ideal laboratory for studying ${\cal N}=4$ SYM. 

We remark that the non-trivial structures in \Roneone amplitudes observed in~\cite{Caron-Huot:2013vda} nicely reflect the algebraic letters and cluster algebra structures of their 4d counterparts. Unlike the $k{+}\ell=2$ case, general \Roneone amplitudes are ``non-factorizable" in the sense that there are terms containing conformal cross-ratios from both even and odd sectors\footnote{For \Roneone kinematics, a null polygon must have an even number of edges, which we denote as $2 n$, and take a zigzag shape where edges with even and odd labels go along the two light-like directions, as we further explain below.}. These non-trivial ``mixing" factors first appear as poles in \Roneone leading singularities/Yangian invariants of one-loop N$^2$MHV octagon, and they emerge as ``mixing letters" at higher loops in general. We will see that, while rational letters of $G_+(4,2n)/T$ reduce to letters of two copies of $G_+(2,n)/T\sim A_{n{-}3}$ (for even and odd sectors), algebraic letters reduce to the mixing letters in \Roneone kinematics. For example, all rational letters of the octagon reduce to $A_1^2:=A_1\times A_1$ letters $v, 1+v$ ($w, 1+w$) with cross-ratios in the odd (even) sector, and algebraic singularities reduce to $v-w$ and $1-v w$. The \Roneone octagon with these $6$ letters is given in terms of the well-known two-dimensional harmonic polylogarithms (2dHPL)~\cite{Gehrmann:2001jv}, or more precisely, two $A_2$ functions with an overlapping $A_1^2$ part, as we will explain shortly (see~\cite{Torres:2013vba} for some earlier results).

Besides amplitudes, we will see even more significant simplifications of certain multi-loop, dual conformally invariant (DCI) Feynman integrals (originally proposed in~\cite{He:2020uxy,He:2021esx}) in \Roneone kinematics, which also support our conjectures about the underlying cluster algebra structure. It is well known that box-ladder integrals can be naturally embedded in \Roneone kinematics where they are given by $A_1^2$ functions. Penta-box ladder integrals, which are $A_3$ functions in 4d, become in \Roneone kinematics the simplest $A_1^2$ functions, ${\rm Li}_\ell(1+v) {\rm Li}_\ell(1+w)$, at $\ell$ loops~\cite{He:2021esx}. We will illustrate our point with more non-trivial examples of eight-point double-pentagon and ladder integrals (similar to those considered in~\cite{Ferro:2012wa}, but with more general numerators) involving square roots which are generally difficult to evaluate in 4d; in \Roneone kinematics the $d\log$ recursion and differential equations they satisfy become much simpler, and it turns out they are all given in terms of $A_2$ functions at most! Moreover, just like amplitudes at $k+\ell=3$~\cite{Caron-Huot:2013vda}, we will show that mixing letters can only appear on the third entry for these integrals, indicating a similar pattern for the algebraic words in 4d.

In addition to this simple alphabet, we propose physical conditions on the first two entries and $\bar Q$-implied conditions on the last two (one) entries for MHV (NMHV) amplitudes, which exclude mixing letters for these entries. Based on these considerations, we will initiate a \Roneone octagon bootstrap program. We will first show how to construct these overlapping $A_2$ functions at any given weight (whose dimension admits a beautiful recursion relation), and find that only a small number of new functions are needed at low weight. We then locate amplitudes with $k+\ell=2,3$, which have been obtained using $\bar{Q}$ equations in~\cite{Caron-Huot:2013vda}, in the cluster function space. We find that $k+\ell=2$ amplitudes can be fixed by dihedral symmetry and collinear limits, while for $k+\ell=3$, we need additionally the soft-collinear operator-product-expansion (OPE) for Wilson loops~\cite{Alday:2010ku,Gaiotto:2010fk,Gaiotto:2011dt,Basso:2013vsa} up to sub-leading orders.

We also briefly comment on the higher-point amplitudes in \Roneone kinematics, which can be uplifted from lower-point building blocks due to strong contraints under collinear-soft limits~\cite{Goddard:2012cx, Caron-Huot:2013vda}. In a precise sense, it has been shown in~\cite{Caron-Huot:2013vda} for $k+\ell\leq 3$ that the ``most complicated part" of higher-point amplitudes can be uplifted from octagons up to a ``remainder" that does not contain mixing letters. This implies that for higher multiplicities with $k+\ell\leq 3$, all we need in \Roneone kinematics are simply $A_2$ functions (one for each square root of four-mass box, with one mixing letter), and $A_{n{-}3}^2$ for the remainder which is free of mixing letters. This is consistent with what we have observed in 4d: algebraic words organize themselves according to square roots of four-mass boxes, with a ``remainder" which contains only rational letters. However, for $k+\ell>3$ more complicated mixing letters, which correspond to more complicated algebraic letters in 4d, may appear.

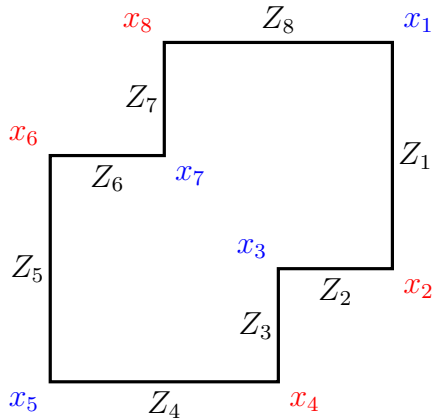
\begin{figure}
    \centering
    \begin{tikzpicture}[scale=1.5]
\draw[black, very thick](2,0)--(2,2)--(0,2)--(0,1)--(-1,1)--(-1,-1)--(1,-1)--(1,0)--cycle;
\filldraw[blue] (2,2) node[anchor=south west] {{$x_1$}};
\filldraw[red] (2,0) node[anchor=north west] {{$x_2$}};
\filldraw[blue] (1,0) node[anchor=south east] {{$x_3$}};
\filldraw[red] (1,-1) node[anchor=north west] {{$x_4$}};
\filldraw[blue] (-1,-1) node[anchor=north east] {{$x_5$}};
\filldraw[red] (-1,1) node[anchor=south east] {{$x_6$}};
\filldraw[blue] (0,1) node[anchor=north west] {{$x_7$}};
\filldraw[red] (0,2) node[anchor=south east] {{$x_8$}};
\node at (1,2.2) {$Z_8$};
\node at (-0.2,1.5) {$Z_7$};
\node at (-0.5,0.8) {$Z_6$};
\node at (-1.2,0) {$Z_5$};
\node at (0,-1.2) {$Z_4$};
\node at (0.8,-0.5) {$Z_3$};
\node at (1.5,-0.2) {$Z_2$};
\node at (2.2,1) {$Z_1$};
\end{tikzpicture}
    \caption{Octagon Wilson loops=$8$-point amplitudes in \Roneone kinematics.}
    \label{2dWL}
\end{figure}

\subsection{Review of \Roneone kinematics}

Let us briefly review some basic facts about \Roneone kinematics~\cite{Caron-Huot:2013vda}. Since the edges of the zigzag-shaped $2n$-gon go along two null directions  $x^{\pm}$  (see Figure~\ref{2dWL} for the octagon), it is convenient to parametrize the momentum twistors~\cite{Hodges:2009hk} as 
\begin{equation}\label{Z2d}
Z_{2i-1}=(\lambda_{2i-1}^1, 0, \lambda_{2i-1}^2,0)\,,\quad Z_{2i}=(0, \tilde\lambda_{2i}^1, 0, \tilde\lambda_{2i}^2),
\end{equation}
which amounts to the reduction $G_+(4, 2n) \to G^{\rm even}_+(2,n) \times G^{\rm odd}_+(2,n)$. Meanwhile, the conformal group $SL(4)$ reduces to $SL(2)\times SL(2)$. The kinematics is encoded in even and odd $SL(2)$-invariants: $\langle i\,j\rangle$ (resp. $[i\,j]$) for odd (resp. even) $i,j$, which are in fact one-dimensional distances along the null direction (in Figure.~\ref{2dWL}, odd/even direction is vertical/horizontal). Any $SL(4)$-invariant expression in four dimensions reduces, and the only four-brakcets that remain non-vanishing are those that involve two odd and two even labels, {\it e.g.} for $i,k$ odd and $j,l$ even  we have $\langle i j k l\rangle = \langle i k\rangle [j l]$.

The most general cross-ratios in \Roneone kinematics are those for the odd and even $A_{n{-}3} \sim G_+(2,n)/T$ cluster algebras: we define $u^{\rm 2d}_{i,j,k,l}:=\frac{[i\,j][k,l]}{[i\,k][j\,l]}$ in the even sector and $[] \to \langle \rangle$ in the odd sector.  Any four-dimensional cross-ratio factorizes into a product of even and odd cross-ratios: for $i,j,k,l$ all of the same parity, the 4d cross-ratio reduces as (our convention is $x_a \sim (a-1 a)$)
\be
u^{4d}_{i,j,k,l}=\frac{x_{ij}^2 x_{kl}^2}{x_{ik}^2 x_{jl}^2} \to u^{\rm 2d}_{i-1, j-1, k-1, l-1} u^{\rm 2d}_{i,j,k,l}\,.
\ee

Amplitudes in \Roneone kinematics are trivial for $2n<8$: for $2n=6$ (hexagon), the all-loop BDS-normalized amplitude $R_{6,0}=R_{6,1}/R_{6,1}^{\rm tree}=1$. 
The first non-trivial case is $2n=8$ (octagon), which corresponds to $A^{\rm even}_1 \times A^{\rm odd}_1$ : using the notation of~\cite{Caron-Huot:2013vda}, we define {\it positive} variables for the odd and even sectors as:
\begin{equation}\label{vw}
v=\frac{u^{\rm 2d}_{1,3,5,7}}{1-u^{\rm 2d}_{1,3,5,7}}=\frac{\langle 1\,3\rangle\langle 5\,7\rangle}{\langle 1\,7\rangle\langle 3\,5\rangle}\in[0,\infty),\quad 
w=\frac{u^{\rm 2d}_{2,4,6,8}}{1-u^{\rm 2d}_{2,4,6,8}}=\frac{[2\,4][6\,8]}{[2\,8][4\,6]}\in[0,\infty), 
\end{equation}
and the letters of these two $A_1$'s are simply $v, 1+v$ and $w, 1+w$. For the $2n$-gon, we have $n(n{-}3)/2$ multiplicatively independent letters for $A^{\rm even}_{n-3}$, and same for $A^{\rm odd}_{n-3}$\footnote{They can be chosen as the {\it dihedral coordinates}~\cite{brown2009multiple} of ${\cal M}_{0,n}$, which satisfy the so-called $u$-equations~\cite{Arkani-Hamed:2017mur,Arkani-Hamed:2019plo}, and they form a multiplicative basis for all possible cross-ratios in that sector. For $2n=8$, the two $u$ variables (which add up to $1$) in the odd sector are, $v/(1+v)$, $1/(1+v)$, and the two in the even sector are $w/(1+w)$, $1/(1+w)$ (all between $0$ and $1$).}. When reduced to \Roneone kinematics, we expect that any cluster variable of $G_+(4,2 n)/T$ simply becomes a monomial of these $n(n{-}3)$ letters. This is the key reason for the simplification in \Roneone kinematics. Starting at $2n=8$ (and $k+\ell>2$), algebraic letters appear in 4d involving square roots of Gram determinants. These, as we will see, reduce to new letters that mix the odd and even sectors.

Next, we quickly review the tree amplitudes and leading singularities (or Yangian invariants) in \Roneone kinematics obtained by supersymmetric reduction from 4d. We reduce the fermionic part of the super-twistors, $\chi_i$'s, in the same way as the bosonic part in \eqref{Z2d}. Thus, the (dual) superconformal group reduces to $SL(2|2)\times SL(2|2)$. The fermionic parts in any Yangian invariant must factorize into two sectors, so the basic $SL(2|2)$-invariants are (for $i,j,k$ odd) 
\be
(i j k):=\frac{\delta^{0\|2}\big(\langle i\,j\rangle
\chi_{k} +\langle j\,k\rangle \chi_{i}+\langle k\,i\rangle \chi_{j}\big)}{\langle i\,j \rangle\langle j\,k\rangle\langle
 k\,i\rangle},\label{Rinv}
\ee
and similarly, $[i j k]$ for even labels with $ \langle \rangle \to [ ] $. Apart from the MHV case (which is by definition $R_{n,0}=1$), the simplest tree amplitude is the NMHV hexagon $R^{\rm tree}_{6,1}=-(1 3 5)[2 4 6]$, and in~\cite{Caron-Huot:2013vda} all tree amplitudes have been obtained as degree-$(k+k)$ polynomials of basic R-invariants with unit coefficients, using a remarkably simple BCFW recursion directly in \Roneone kinematics. 

Note that in \Roneone kinematics the highest $k$ possible is $k=n-2$, in which case the tree amplitude takes the simplest form: it is given by the product of $n{-}2$ even R-invariants and $n{-}2$ odd ones, which has a beautiful geometric interpretation. Replacing $\chi_a$ by $d Z_a$~\cite{Arkani-Hamed:2017vfh,He:2018okq} in the basic R-invariants $(ijk)$, it becomes the canonical form $(i j k) \to d\log \frac{\langle i j\rangle }{\langle i k\rangle} d\log \frac{\langle j k \rangle}{\langle i k\rangle}$ of a triangle with vertices $Z_i, Z_j, Z_k$. For R-invariants of higher $k$, we have the $k$-th power of the $2$-form of an $n$-gon, which is a well-known result of the $m=2$ tree amplituhedron~\cite{Arkani-Hamed:2013jha}. For example, for $2 n=8, k=2$ we have
\begin{equation}
R^{\rm tree}_{8,2}=(135)(357)[246][468]=\left(\frac{(135)+(357)}2\right)^2 \left(\frac{[246]+[468]}2\right)^2=:(1 3 5 7)[2 4 6 8].
\end{equation}

Leading singularities, on the other hand, usually do not have smooth 2d limits (although any complete amplitude/WL must have), and it is still an important open question how to compute them directly in \Roneone kinematics\footnote{One can define the \Roneone amplituhedron by reducing external data to \Roneone kinematics, but individual Yangian invariants generally do not have smooth limits. Many $4k$-dim positroid cells, which give leading singularities/Yangian invariants in 4d, do not have well-defined 2d limits.}. Here, we content ourselves by simply summarizing the results for octagons. For NMHV, we have $3$ independent R-invariants in each sector: $(1):=(1 3 5)$ and its $3$ cyclic rotations in the odd sector satisfy $(1)-(3)+(5)-(7)=0$, and similarly for $[2]:=[2 4 6]$ {\it etc.} in the even sector with $[2]-[4]+[6]-[8]=0$. We can rewrite the $3\times 3=8+1$ independent Yangian invariants using the length-$8$ cyclic orbit of $(1)[2]$ together with $R_{8,1}^{\rm tree}=(3)[6]-(3)[4]-(1)[8]-(5)[6]$. Accompanying them are two independent functions of $v,w$, so that to all loops we have
\begin{equation}
R_{8,1}=\big( \big([2](3) +[6](7)\big)f_{8,1}^1(v, w) + \textrm{3 cyclic} \big) + R_{8,1}^\textrm{tree} f_{8,1}^2(v,w). \label{R81}
\end{equation}

For N$^2$MHV octagons, since all Yangian invariants are proportional to $R^{\rm tree}_{8,2}$ above, they differ by possible prefactors (functions of $v,w$) which {\it mix} the odd and even sectors. A general argument based on positroid cells of $G_+(4,8)$ concludes that there are only $4$ pre-factors:
\begin{equation}
\frac{v}{v-w}\,,\quad \frac{v w}{1- v w}\,,\quad \frac{w}{v-w}\,,\quad \frac{1}{1- v w}\,,
\end{equation}
which (up to a sign) are related by cyclic rotations. The non-trivial linear combinations $\frac{v+w}{v-w}$ and $\frac{1+vw}{1-vw}$ give exactly the coefficients of the two four-mass-boxes in \Roneone. To all loops, no other leading singularities can appear for N$^2$MHV octagon. We write 
\be~\label{R82}
R_{8,2}=R_{8,2}^{\rm tree} \left(\frac{v}{v-w} F_{8,2} (v,w) + {\rm cyclic} \right)\,.
\ee

\section{Octagons: two overlapping $A_2$'s and cluster bootstrap}

In this section we consider the main object of interest, the octagon ($2 n=8$): we provide strong evidence for its remarkably simple alphabet, which can be used to efficiently bootstrap and constrain higher-loop amplitudes and Feynman integrals. It has already been observed in~\cite{Caron-Huot:2013vda} that only $6$ letters, $v, 1+v, w, 1+w, v-w, 1-v w$ appear for $k+\ell=3$; the first four are from $A_1^2$, and the last two mix the even and odd sectors. It is easy to see that these mixing letters are \Roneone avatars of algebraic singularities in 4d containing Gram-determinant square roots for four-mass boxes. For octagons, the two mixing letters correspond respectively to the two four-mass configurations $(x_1, x_3, x_5, x_7)$ and $(x_2, x_4, x_6, x_8)$ (these vertices are labelled by blue and red $x$'s respectively in Figure~\ref{2dWL}). More precisely, with $U:=u^{\rm 4d}_{a,b,c,d}$ and $V:=u^{\rm 4d}_{b,c,d,a}$ for $(x_a, x_b, x_c, x_d)$, in \Roneone kinematics the square root $\Delta_{a,b,c,d}:=\sqrt{(1+U-V)^2-4 U V}$ reduces to\footnote{We label any four-mass box by the $4$ off-shell dual points, $x_a, x_b, x_c, x_d$, which differ from the notation in~\cite{ArkaniHamed:2010gh} and~\cite{Caron-Huot:2013vda}: {\it e.g.} for the octagon, the $2$ four-mass boxes are swapped.}
\begin{equation}\label{mixing}
\Delta_{2,4,6,8}=\frac{|v-w|}{(1+v)(1+w)}\,,\quad \Delta_{1,3,5,7}=\frac{|1-v w |}{(1+v)(1+w)}\,.
\end{equation}
Equation \eqref{R82} is a preview of these mixing letters as non-trivial poles for (the leading singularities of) one-loop N$^2$MHV; they really start to appear in the third symbol-entry of two-loop NMHV and three-loop MHV.

Note that the six letters form a $C_2$ cluster algebra: with $z_1=v$, $z_2=-v/w$, the alphabet becomes $\{z_1, 1+z_1, z_2, 1+z_2, z_1-z_2, z_1^2+z_2\}$, which is in accordance with the notation of~\cite{Chicherin:2020umh}. Thus as we mentioned, all octagon functions belong to the class of 2dHPL. However, octagon functions are very special 2dHPL since the two mixing letters never appear in the same term, which follows from the fact that in 4d the two square roots cannot appear together. For this reason, we will refer to the alphabet as two overlapping $A_2$ 's (rather than a $C_2$): one with $v, 1+v, w, 1+w, v-w$ and the other with $1-v w$ instead. The second $A_2$ (associated with $\Delta_{1,3,5,7}$) is related to the first one (associated with $\Delta_{2,4,6,8}$) by a cyclic rotation $i\to i{-}1$, or equivalently $v \to 1/w$ and $w\to v$, so that the letters become $1/w, 1+1/w, v, 1+v, 1/w-v=(1-v w)/w$.

Now we state our main {\bf conjecture} of the section: {\it To all loops, the octagon $R_{8,k}$ with $k=0,1,2$ can be written as linear combinations of two $A_2$ cluster functions, sharing the letters $v,w,1+v, 1+w$ but one has $v-w$ in addition and the other has $1- v w$, and the coefficients are given by \Roneone leading singularities.} 

As a refinement, we further conjecture that {\it the first two entries and last two entries for MHV (and last entry for NMHV) are free of mixing letters, {\it i.e.} they are from $A_1^2$.} The first-two-entry conditions follow from physical discontinuity (for the first entry) and Steinmann relations~\cite{Caron-Huot:2016owq}, so that only four-mass boxes (containing ${\rm Li}_2$) and degenerations (only logarithmic functions) are allowed in the first two entries.
Using $\bar Q$ equations in \Roneone kinematics~\cite{Caron-Huot:2013vda}, the last two entries for MHV are determined by leading singularities and last entries of NMHV decagon, which are in turn determined by leading singularities of N$^2$MHV amplitudes. These derivations are independent of loop orders, and we find the last two entries for MHV and the last one entry for NMHV are all free of mixing letters. 

Before providing explicit results of octagon amplitudes/integrals which support our conjecture, let us give some general arguments and explanations. First, the $A_1^2$ part represents the \Roneone reduction of all possible cluster variables of $G_+(4,8)/T$. We have explicitly evaluated the $440$ letters provided in~\cite{Henke:2019hve}, and find that in \Roneone kinematics they all become monomials of $v, w, 1+v, 1+w$. 

Furthermore, for either four-mass box, $(x_2,x_4,x_6,x_8)$ $(x_1,x_3,x_5,x_7)$ of the octagon, {\it any} algebraic letter can be written as ${\bf x}-\alpha_{\pm}$ with the two roots satisfying $\alpha_+ \alpha_-=U$, $(1-\alpha_+)(1-\alpha_-)=V$, where ${\bf x}$ is a DCI, rational function of Pl{\"u}cker coordinates. Our conjecture states that this combination must reduce to either $v-w$ or $1-v w$ (up to prefactors that are monomials of letters in $A_1^2$). There are two cases: for the four-mass boxes appeared in one-loop N$^2$MHV, {\it e.g.} $F(2,4,6,8)$, we have ``trivial" algebraic letters with ${\bf x}=0$ or $1$, which do not lead to any mixing (still in $A_1^2$); the symbol ${\cal S}[F(2,4,6,8)]$ reduces to~\cite{Caron-Huot:2013vda}:
\begin{equation}\label{4m2468}
\frac 1 2 \bigg( U\otimes \frac{1-\alpha_-}{1-\alpha_+}+V \otimes \frac{\alpha_+}{\alpha_-}\bigg) =\frac 1 2 \bigg( v w \otimes \frac{1+v}{1+w}-(1+v)(1+w)\otimes \frac{v}{w}\bigg) \,.
\end{equation}
For non-trivial algebraic letters, it is remarkable that in \Roneone kinematics the rational function ${\bf x}$ always degenerates to $\alpha_+$ or $\alpha_-$, thus the only non-trivial combination is $\alpha_+-\alpha_-=\Delta$, exactly producing the two mixing letters in \eqref{mixing}. We have explicitly checked that this is true for the $9+9$ independent algebraic letters, first computed for the two-loop NMHV octagon~\cite{Zhang:2019vnm}. 

Let us quickly review octagons with $k+\ell\leq 3$ computed using $\bar Q$ equations in~\cite{Caron-Huot:2013vda}. For $k+\ell=2$, {\it i.e.} one-loop NMHV and two-loop MHV, only $A_1^2=\{v, 1+v, w, 1+w\}$ appear (and they are logarithmic functions of weight $2$ and $4$ respectively), which is consistent with the fact that they have no algebraic singularities. For $k+\ell=3$, the first appearance of algebraic singularities is in the pre-factor of one-loop N$^2$MHV: in front of $F(2,4,6,8)$ we have a pole at  $\Delta_{2,4,6,8} \propto v{-}w=0$, and for $F(1,3,5,7)$ a pole at $\Delta_{1,3,5,7} \propto 1{-}v w=0$; this is consistent with the fact that as $v-w$ (or $1-v w$) flips sign, so does $F(2,4,6,8)$ (or $F(1,3,5,7)$). Remarkably, these prefactors become mixing letters for higher-loop amplitudes. For both two-loop NMHV and three-loop MHV, the complete results are recorded in~\cite{Caron-Huot:2013vda}: for the former, it is worth noticing that at while $f^2_{8,1}$ is an $A_1^2$ function, $f^1_{8,1}$ is given by the sum of two $A_2$ functions, and the same is true for $R_{8,0}^{\rm 3-loop}$. 

Furthermore, it has been noted in appendix D of~\cite{Caron-Huot:2013vda} that mixing letters only appear on the third entry. For $f^2_{8,1}$ this is implied by our refined conjecture that first two entires and last entry are free of them, and for $R_{8,0}^{\rm 3-loop}$ the situation is even better since they are also absent in the fourth entry. In both cases, the first two entries in front of $v-w$ is exactly the symbol of $F(2,4,6,8)$ in \eqref{4m2468} (similarly for $1-v w$ we find that of $F(1,3,5,7)$); these can be seen by taking $(2,1,1)$- and $(2,1,3)$-coproducts respectively. The former is fully consistent with our results in 4d~\cite{Zhang:2019vnm} since non-trivial algebraic letters only appear on the third entry (accompanied by four-mass boxes in the first two entries). The latter provides a very important prediction for $R_{8,0}^{\rm 3-loop}$ in 4d, namely non-trivial algebraic letters can only appear on the third entry as well!

In general, we expect that by using $\bar{Q}$ equations, one can predict such mixing letters for MHV and NMHV amplitudes from NMHV and N$^2$MHV amplitudes at one-lower loop order. The derivation from two to three loops in~\cite{Caron-Huot:2013vda} could be formulated more generally for higher loops if we assume the alphabet of NMHV decagon (as we will explain in the next section). This argument should provide a proof that only $v-w$ and $1- v w$ can appear for higher loop octagons. For now, we content ourselves with explicit results for amplitudes (up to $k+\ell=3$) and a series of two-loop and all-loop Feynman integrals.  


\subsection{Octagon integrals from two overlapping $A_2$'s}

Although amplitudes/WL up to $k+\ell=3$ provides strong evidence for our conjecture, since currently we have no analytic access to higher loops, it is desirable to see if we have more examples supporting our conjecture. This is why we turn to studying Feynman integrals, including all finite double-pentagon integrals for two-loop MHV and NMHV amplitudes, as well as certain ladder integrals to all loops. Their evaluations strongly support our conjecture: all of them are given in terms of two overlapping $A_2$ functions or even simpler functions {\it e.g.} one $A_2$ function, $A_1^2$ function or even just a single $A_1$ function. Moreover, they are constrained by such cluster algebra structures so strongly that we find large ``degeneracy" in functions they evaluate to. 

\paragraph{One Loop}Let us begin by considering the one-loop case. It turns out we have exactly two kinds of weight-2 functions (linear combinations of two overlapping $A_2$ functions, having only physical discontinuities and satisfying Steinmann relations), all of them $A_1^2$ functions: two four-mass boxes and 10 logarithmic functions. We record the four-mass box $F(2,4,6,8)$ (the other four-mass box $F(1,3,5,7)$ is its cyclic image) whose symbol is given in \eqref{4m2468}:
\be
F(2,4,6,8)=L(w)-L(v)+\frac12\log(v)\log(1{+}w)-\frac12\log(w)\log(1{+}v),\quad\text{for }v>w,
\ee 
with $v \leftrightarrow w$ if $v<w$, where $L(x):={\rm Li}_2(-x)+\frac 12\log(x)\log(1{+}x)+\frac{\pi^2}{12}$ is the Rogers dilogarithm.

Let us take a look at some examples. Consider the two natural chiral pentagons needed for MHV and NMHV amplitudes with different numerators~\cite{ArkaniHamed:2010gh}, which are clearly different from four-mass boxes in 4d,
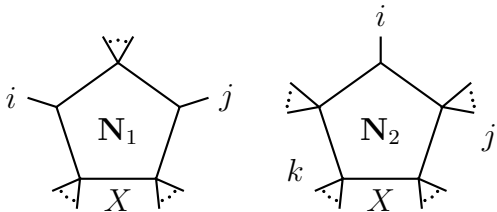
\vspace{-2ex}
\[\begin{tikzpicture}[baseline={([yshift=-.5ex]current bounding box.center)},scale=0.2]
        \draw[black,thick] (0,0)--(5,0)--(6.55,4.76)--(2.50,7.69)--(-1.55,4.76)--cycle;
        \draw[black,thick] (1.5,9.43)--(2.5,7.69)--(3.5,9.43);
        \filldraw[black] (2.5,9.19) circle [radius=2pt];
        \filldraw[black] (1.99,9.1) circle [radius=2pt];
        \filldraw[black] (3.01,9.1) circle [radius=2pt];
        \draw[black,thick] (-0.21,-1.99)--(0,0)--(-1.83,-0.81);
        \filldraw[black] (-0.88,-1.21) circle [radius=2pt];
        \filldraw[black] (-0.41,-1.44) circle [radius=2pt];
        \filldraw[black] (-1.24,-0.84) circle [radius=2pt];
         \draw[black,thick] (6.55,4.76)--(8.45,5.37);
        \filldraw[black] (5.88,-1.21) circle [radius=2pt];
        \filldraw[black] (6.24,-0.84) circle [radius=2pt];
        \filldraw[black] (5.41,-1.44) circle [radius=2pt];
        \draw[black,thick] (-1.55,4.76)--(-3.45,5.37);
         \draw[black,thick] (6.83,-0.81)--(5,0)--(5.21,-1.99);
        \filldraw[black] (-3.45,5.37) node[anchor=east] {{$i$}};
        \filldraw[black] (8.45,5.37) node[anchor=west] {{$j$}};
        \filldraw[black] (2.5,0) node[anchor=north] {{$X$}};
         \filldraw[black] (2.5,4.76) node[anchor=north] {{${\bf N}_1$}};
    \end{tikzpicture}\quad
\begin{tikzpicture}[baseline={([yshift=-2ex]current bounding box.center)},scale=0.2]
        \draw[black,thick] (0,0)--(5,0)--(6.55,4.76)--(2.50,7.69)--(-1.55,4.76)--cycle;
        \draw[black,thick] (2.5,7.69)--(2.5,9.43);
        \draw[black,thick] (-0.21,-1.99)--(0,0)--(-1.83,-0.81);
        \filldraw[black] (-0.88,-1.21) circle [radius=2pt];
        \filldraw[black] (-0.41,-1.44) circle [radius=2pt];
        \filldraw[black] (-1.24,-0.84) circle [radius=2pt];
        \draw[black,thick] (6.83,-0.81)--(5,0)--(5.21,-1.99);
        \filldraw[black] (5.88,-1.21) circle [radius=2pt];
        \filldraw[black] (6.24,-0.84) circle [radius=2pt];
        \filldraw[black] (5.41,-1.44) circle [radius=2pt];
        \filldraw[black] (8.45,5.7) circle [radius=2pt];
        \filldraw[black] (8.55,5.2) circle [radius=2pt];
        \filldraw[black] (8.65,4.8) circle [radius=2pt];
        \filldraw[black] (-3.45,5.7) circle [radius=2pt];
        \filldraw[black] (-3.55,5.2) circle [radius=2pt];
        \filldraw[black] (-3.65,4.8) circle [radius=2pt];
         \draw[black,thick] (8.65,4.37)--(6.55,4.76)--(8.45,6.37);
        \draw[black,thick] (-3.45,6.37)--(-1.55,4.76)--(-3.65,4.37);
        \filldraw[black] (2.5,9.43) node[anchor=south] {{$i$}};
        \filldraw[black] (8.45,4.37) node[anchor=north west] {{$j$}};
        \filldraw[black] (-1.83,-0.9) node[anchor=south east] {{$k$}};
        \filldraw[black] (2.5,0) node[anchor=north] {{$X$}};
        \filldraw[black] (2.5,4.76) node[anchor=north] {{${\bf N}_2$}};
    \end{tikzpicture}
\]
\[{\bf N}_1=\l \ell \bar{i} \cap \bar{j}\r,\ {\bf N}_2=\frac12\l\ell\bar i\cap(jj{+}1((ikk{+}1)\cap X))-(kk{+}1((ijj{+}1)\cap X))\r.\]

For octagon \Roneone kinematics, we can choose $i=1$, $j=4$, and $X=(67)$ in the first case, and $i=1$, $j=2$ and $X=(45)$ in the second case. Remarkably, they both reduce to the above two kinds of functions: the first pentagon integral evaluates to $I_{\rm pen.}=\log(1+v)\log (1+w)$, while the second one becomes $F(1,3,5,7)$. The fact that different integrals in 4d reduces to the same overlapping $A_2$ functions in \Roneone illustrates how constraining such cluster algebra structures are. It is also interesting that although in 4d both type of pentagons contribute to one-loop NMHV ratio functions, the final result in \Roneone contains logarithmic functions only (four-mass boxes, or ${\rm Li}_2$ functions, cancel out and only start to contribute for $k+\ell\geq2$ amplitudes).

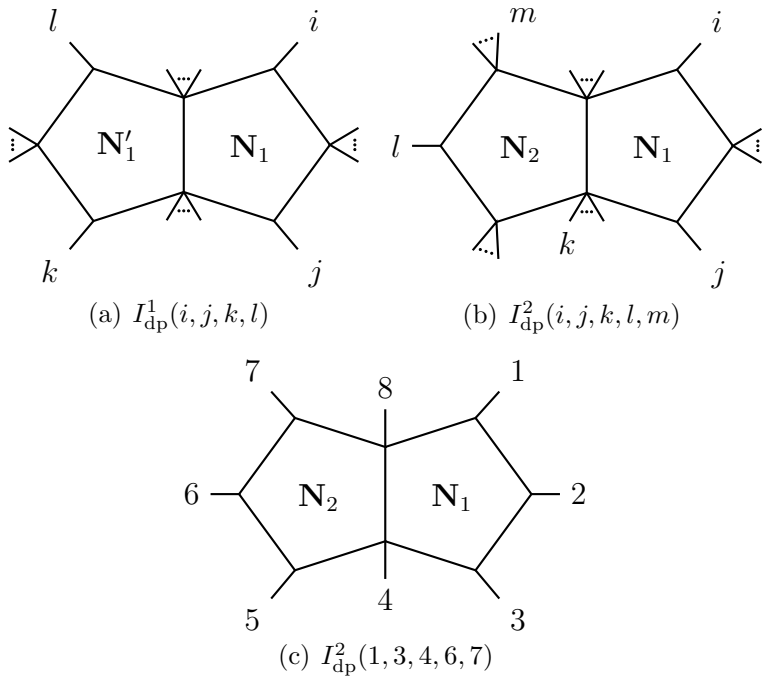
\begin{figure}[htbp]
\centering
\subfigure[$I_{\rm dp}^1(i,j,k,l)$]{
\begin{tikzpicture}[baseline={([yshift=-.5ex]current bounding box.center)},scale=0.25]
\draw[black,thick](0,0)--(0,5)--(4.75,6.54)--(7.69,2.50)--(4.75,-1.54)--cycle;
\draw[black,thick](0,5)--(-4.75,6.54)--(-7.69,2.50)--(-4.75,-1.54)--(0,0);
\draw[black,thick](-0.9,6.5)--(0,5)--(0.9,6.5);
\filldraw[black]  (0,6) circle [radius=1.5pt];
\filldraw[black]  (-0.25,6) circle [radius=1.5pt];
\filldraw[black]  (0.25,6) circle [radius=1.5pt];
\draw[black,thick](-0.9,-1.5)--(0,0)--(0.9,-1.5);
\filldraw[black]  (0,-1) circle [radius=1.5pt];
\filldraw[black]  (-0.25,-1) circle [radius=1.5pt];
\filldraw[black]  (0.25,-1) circle [radius=1.5pt];
\draw[black,thick](4.75,6.54)--(6,7.94);
\draw[black,thick](4.75,-1.54)--(6,-3.04);
\draw[black,thick](9.19,1.6)--(7.69,2.50)--(9.19,3.4);
\draw[black,thick](-4.75,6.54)--(-6,7.94);
\draw[black,thick](-4.75,-1.54)--(-6,-3.04);
\draw[black,thick](-9.19,1.6)--(-7.69,2.50)--(-9.19,3.4);
\filldraw[black]  (9,2.5) circle [radius=1.5pt];
\filldraw[black]  (9,2.25) circle [radius=1.5pt];
\filldraw[black]  (9,2.75) circle [radius=1.5pt];
\filldraw[black]  (-9,2.5) circle [radius=1.5pt];
\filldraw[black]  (-9,2.25) circle [radius=1.5pt];
\filldraw[black]  (-9,2.75) circle [radius=1.5pt];
\filldraw[black] (6,7.94) node[anchor=south west] {{$i$}};
\filldraw[black] (6,-3.04) node[anchor=north west] {{$j$}};
\filldraw[black] (-6,7.94) node[anchor=south east] {{$l$}};
\filldraw[black] (-6,-3.04) node[anchor=north east] {{$k$}};
\filldraw[black] (3.5,1.04) node[anchor=south] {{${\bf N}_1$}};
\filldraw[black] (-3.5,1.04) node[anchor=south] {{${\bf N}_1^\prime$}};
\end{tikzpicture}}
\subfigure[$I_{\rm dp}^2(i,j,k,l,m)$]{
\begin{tikzpicture}[baseline={([yshift=-.5ex]current bounding box.center)},scale=0.25]
\draw[black,thick](0,0)--(0,5)--(4.75,6.54)--(7.69,2.50)--(4.75,-1.54)--cycle;
\draw[black,thick](0,5)--(-4.75,6.54)--(-7.69,2.50)--(-4.75,-1.54)--(0,0);
\draw[black,thick](-0.9,6.5)--(0,5)--(0.9,6.5);
\filldraw[black]  (0,6) circle [radius=1.5pt];
\filldraw[black]  (-0.25,6) circle [radius=1.5pt];
\filldraw[black]  (0.25,6) circle [radius=1.5pt];
\draw[black,thick](-0.9,-1.5)--(0,0)--(0.9,-1.5);
\filldraw[black]  (0,-1) circle [radius=1.5pt];
\filldraw[black]  (-0.25,-1) circle [radius=1.5pt];
\filldraw[black]  (0.25,-1) circle [radius=1.5pt];
\draw[black,thick](4.75,6.54)--(6,7.94);
\draw[black,thick](4.75,-1.54)--(6,-3.04);
\draw[black,thick](9.19,1.6)--(7.69,2.50)--(9.19,3.4);
\draw[black,thick] (-4.65,8.50)--(-4.75,6.54)--(-6,7.94);
\draw[black,thick] (-4.65,-3.50)--(-4.75,-1.54)--(-6,-3.04);
\draw[black,thick](-9.19,2.5)--(-7.69,2.50);
\filldraw[black]  (9,2.5) circle [radius=1.5pt];
\filldraw[black]  (9,2.25) circle [radius=1.5pt];
\filldraw[black]  (9,2.75) circle [radius=1.5pt];
\filldraw[black] (6,7.94) node[anchor=south west] {{$i$}};
\filldraw[black] (6,-3.04) node[anchor=north west] {{$j$}};
\filldraw[black] (-4.65,8.50) node[anchor=south west] {{$m$}};
\filldraw[black] (-9.19,2.5) node[anchor=east] {{$l$}};
\filldraw[black] (0,-1.5) node[anchor=north east] {{$k$}};
\filldraw[black] (3.5,1.04) node[anchor=south] {{${\bf N}_1$}};
\filldraw[black] (-3.5,1.04) node[anchor=south] {{${\bf N}_2$}};
\filldraw[black]  (-5,8.2) circle [radius=1.5pt];
\filldraw[black]  (-5.3,8.1) circle [radius=1.5pt];
\filldraw[black]  (-5.6,8) circle [radius=1.5pt];
\filldraw[black]  (-5,-3.2) circle [radius=1.5pt];
\filldraw[black]  (-5.3,-3.1) circle [radius=1.5pt];
\filldraw[black]  (-5.6,-3) circle [radius=1.5pt];
\end{tikzpicture}
}
\subfigure[$I_{\rm dp}^2(1,3,4,6,7)$]{\begin{tikzpicture}[baseline={([yshift=-.5ex]current bounding box.center)},scale=0.25]
\draw[black,thick](0,0)--(0,5)--(4.75,6.54)--(7.69,2.50)--(4.75,-1.54)--cycle;
\draw[black,thick](0,5)--(-4.75,6.54)--(-7.69,2.50)--(-4.75,-1.54)--(0,0);
\draw[black,thick](0,7)--(0,5);
\draw[black,thick](0,-2)--(0,0);
\draw[black,thick](4.75,6.54)--(6,7.94);
\draw[black,thick](4.75,-1.54)--(6,-3.04);
\draw[black,thick](9.19,2.5)--(7.69,2.50);
\draw[black,thick] (-4.75,6.54)--(-6,7.94);
\draw[black,thick] (-4.75,-1.54)--(-6,-3.04);
\draw[black,thick](-9.19,2.5)--(-7.69,2.50);
\filldraw[black] (6,7.94) node[anchor=south west] {{$1$}};
\filldraw[black] (9.19,2.5) node[anchor=west] {{$2$}};
\filldraw[black] (6,-3.04) node[anchor=north west] {{$3$}};
\filldraw[black] (-6,-3.04) node[anchor=north east] {{$5$}};
\filldraw[black] (-6,7.94) node[anchor=south east] {{$7$}};
\filldraw[black] (0,-2) node[anchor=north] {{$4$}};
\filldraw[black] (0,7) node[anchor=south] {{$8$}};
\filldraw[black] (-9.19,2.5) node[anchor=east] {{$6$}};
\filldraw[black] (3.5,1.04) node[anchor=south] {{${\bf N}_1$}};
\filldraw[black] (-3.5,1.04) node[anchor=south] {{${\bf N}_2$}};
\end{tikzpicture}}
\caption{Double pentagon integrals}\label{fig:dp}
\end{figure}

\paragraph{Two Loops}Now we move to two loops. Naively there are more than one hundred different weight-$4$ functions (see below), but we restrict ourselves to double pentagon integrals needed for MHV and NMHV amplitudes, where we have certain numerators for both loops (see~\cite{ArkaniHamed:2010gh,Bourjaily:2018aeq} for details), as shown in Figure~\ref{fig:dp}.

We use ${\bf N}_r$ for $r=1,2$ to denote the two types of numerators for these pentagons. For double pentagons needed for MHV amplitudes, $I^1_{\rm dp}(i,j,k,l)$ with ${\bf N}_1=\l \ell \bar{i} \cap \bar{j}\r$ and ${\bf N}_1^\prime=\l \ell' \bar{k} \cap \bar{l}\r$,
there are five inequivalent topologies which are finite for $2n=8$: $(i,j,k,l)= (1,3,4,7) , (1,3,4,8),(1,3,5,7), (1,3,6,8)$ and $(1,4,5,8)$. It is clear that (1,3,4,7) and (1,3,5,7) vanish in \Roneone (even without putting the overall numerator $\l 1347\r=\l 1357\r=0$), and the rest have been evaluated in~\cite{He:2020lcu}. We find that $(1,3,4,8)$ is given by a simple $A_1$ function, while $(1,3,6,8)$ and $(1,4,5,8)$ are both $A_2$ functions. The reason why $I^1_{\rm dp}(1,3,4,8)$ is only $A_1$ is that, since it does not depend on the dual point $x_6$, only three momentum twistors with even labels can appear in the final answer, which are insufficient to form a cross ratio, so that $I^1_{\rm dp}(1,3,4,8)$ only depends on $v$.  For the two $A_2$ functions, explicit computations show that their $(2,1,1)$-coproducts are:
\begin{equation}
    \Delta_{2,1,1}(I^1_{\rm dp}(1,4,5,8))=F(2,4,6,8) \otimes(v{-}w)\otimes\frac vw+{\rm non{-}mixing\ part}
\end{equation}
\begin{equation}
    \Delta_{2,1,1}(I^1_{\rm dp}(1,3,6,8))=F(2,4,6,8)\otimes(v{-}w)\otimes\frac {1{+}v}{1{+}w}+{\rm non{-}mixing\ part}
\end{equation}
which support our conjecture since the two integrals both have four off-shell dual points $\{x_2,x_4,x_6,x_8\}$. After a cyclic rotation, $w\to \frac1v$ and $v\to w$, $\{x_2,x_4,x_6,x_8\}$ turn into $\{x_1,x_3,x_5,x_7\}$ and the mixing letter $v{-}w$ becomes $1{-}v w$ as we expected. Furthermore, after two cyclic rotations, $v\to \frac1v$ and $w\to\frac1w$, the mixing part of $ \Delta_{2,1,1}(I^1_{\rm dp}(1,4,5,8))$ comes back to itself, while the mixing part of  $\Delta_{2,1,1}(I^1_{\rm dp}(1,3,6,8))$ becomes that of $\Delta_{2,1,1}(I^1_{\rm dp}(1,3,6,8))+\Delta_{2,1,1}(I^1_{\rm dp}(1,4,5,8))$.

For another type of double pentagons $I^2_{\rm dp}(i,j,k,l,m)$ with numerators ${\bf N}_1=\l \ell \bar{i} \cap \bar{j}\r$ and ${\bf N}_2=\frac12\l\ell^\prime\bar l\cap(mm{+}1((lkk{+}1)\cap (ij)))-(kk{+}1((lmm{+}1)\cap (ij)))\r$ (Figure~\ref{fig:dp}(b)), we have $22$ different integrals, $13$ of which are non-trivial in \Roneone kinematics. Note that as ${\bf N}_2$ greatly simplifies in \Roneone kinematics, some of these integrals evaluate to the same results as the previous integrals. For instance, $I^2_{\rm dp}(1,3,3,6,8)$ is equal to a penta-box integral. 
One nontrivial integral is Figure~\ref{fig:dp} (c).  Contrary to $I^1_{\rm dp}(1,3,5,7)$, now the numerator ${\bf N}_2\propto \l\ell^\prime57\r$ in \Roneone, which gives a non-vanishing result. This integral depends on all $8$ dual points $x_1\sim x_8$, indicating that its symbol has both mixing letters. Explicit computation shows 
\begin{multline}
    \Delta_{2,1,1}(I^2_{\rm dp}(1,3,4,6,7))\\
    =F(2,4,6,8) \otimes(v{-}w)\otimes\frac {1{+}v}{1{+}w}+\left(v\to\frac1w,\ w\to v\right)+\text{non-mixing part},
\end{multline}
where $(v\to\frac1w,\ w\to v)$ means a cyclical rotation in the reverse direction. Indeed, the integral is a cluster function with two overlapping $A_2$'s, as we would expect for the most general octagon in \Roneone.

\paragraph{All-loop Ladders}Finally we move to some all-loop integrals, and the simplest case with algebraic singularity is the double-penta-ladder which we denote as $I^{(L)}(1,4,5,8)$; for $L=2$ it reduces to the double-pentagon integral $I_{\rm dp}^1(1,4,5,8)$ but one can naturally extend it to all loops by inserting a box ladder in between. The integral is expected to contain only one square root $\Delta_{2,4,6,8}$, and in 4d the calculation based on $d\log$ recursion is quite involved due to the need of rationalization (no result beyond $L=2$ has been obtained so far). We show that in \Roneone, the $d\log$ recursion becomes trivial to solve, and the integral evaluates to a single $A_2$ function with the mixing letter $v-w$ only appearing on the third entry. 

By reducing the result of~\cite{He:2020uxy} to \Roneone or directly write down $d\log$ forms, we see that the $d\log$ recursion takes the remarkably simple form:
\be
I^{(L+1)}(v,w)=\int_{1}^\infty d\log t\int_{1}^\infty d\log s\, I^{(L)}(vt,ws),
\ee
where the source of the recursion is the one-loop hexagon:
\be I^{(1)}=\frac{v}{v-w}J(v,w)+\frac{w}{w-v}J(w,v),\ee 
with $J(v,w):=-2\,\Li_2\left(\frac1{1+v}\right)+\log\left(\frac{v}{1+v}\right)\log(w(1+v))$. 
We see that, similarly to the $k+\ell=3$ amplitude, beautifully the mixing letter $v-w$ first appears as a prefactor of the one-loop result, which manifests the symmetry under $v\leftrightarrow w$. Conversely, if we rewrite the recursion as
\[
I^{(L+1)}(v,w)=\int_{v}^\infty d\log t\int_{w}^\infty d\log s\, I^{(L)}(t,s),
\]
and it is easy to get a beautiful second-order differential equation
\begin{equation}
I^{(L)}=v\partial_v w\partial_w I^{(L+1)}.
\end{equation}
The $d\log$ recursion, or differential equation, allows us to directly compute $I^{(L)}(1,4,5,8)$ to all loops (including odd-weight objects in between). From weight $2$ to $3$, one can absorb the prefactor to get a more non-trivial $d\log$ form (analog of the rationalization we did in 4d), and starting from there we always have pure functions, which trivializes any further rationalization encountered for higher loops in 4d.

This computation reveals a remarkable phenomenon for the algebraic letters generalizing what we have seen at two loops: the mixing letter $v-w$ appears {\it only} on the third entry, with the first two accompanying it given by the symbol of four-mass box, and subsequent ones only containing $v/w$ (recall that the alphabet of $I^{(L)}$ is
$\{v, 1 + v, w, 1 + w, v - w\}$). Let us first show that in the last two entries, only $v$ and $w$ are allowed. We note that 
\[
v\partial_v I^{(L+1)}=-\int_w^\infty d\log s\, I^{(L)}(v,s)\quad \text{and}\quad w\partial_w I^{(L+1)}=-\int_v^\infty d\log t \, I^{(L)}(t,w)
\]
are pure functions for $L\geq 2$. Now suppose the symbol of $I^{(L)}$ $(L\geq 2)$ takes the form
\[
    \mathcal S(I^{(L)})=S^{(L)}_w\otimes w+S^{(L)}_v\otimes v+S^{(L)}_{1+w}\otimes (1+w)+S^{(L)}_{1+v}\otimes (1+v)+S^{(L)}_{v-w}\otimes (v-w),
\]
and we obtain
\[
w\partial_w I^{(L)}=S^{(L)}_w+\frac{w}{1+w}S^{(L)}_{1+w}-\frac{w}{v-w}S^{(L)}_{v-w}.
\]
Since $w\partial_w I^{(L)}$ is pure, $S^{(L)}_{1+w}$ and $S^{(L)}_{v-w}$ must vanish; symmetrically, $S^{(L)}_{1+v}$ vanishes because $v\partial_v I^{(L)}$ is pure. Therefore, we have proven that the last entries of $I^{(L)}$ for $L\geq 2$ can only be $v$ or $w$. In fact, one can further prove that for $L>2$, only $v$ and $w$ occur at the last two entries.

For the mixing part with $v-w$ on the third entry, by the symmetry between $v$ and $w$ and $\mathcal S(F(v,w))=-\mathcal S(F(w,v))$ for four-mass box $F(2,4,6,8)=F(v,w)$, the last entry for $L=2$ can only be $v/w$. We will not give the detail of the proof here, but after some algebra one can show that by further integrations from $L=2$, the mixing part of $\Delta_{2,1, 2L-3} (I^{(L)})$ for $L\geq 2$ takes the remarkably simple form:
\begin{equation}
F(v,w)\otimes (v-w)\otimes \frac{1}{(2L{-}3)!} \log^{2L-3} \left(\frac{v}{w} \right).
\end{equation}

\subsection{Cluster bootstrap for octagons}

Based on the alphabet, we are ready to consider the cluster bootstrap for octagons. We emphasize that these octagon amplitudes in \Roneone kinematics with $k{+}l\leq 3$ have been obtained in~\cite{Caron-Huot:2013vda} so our results here are not new. However, our main point is to illustrate how simple the relevant octagon space has become and, even in the absence of more OPE data, one can locate these amplitudes in the space with relative ease. We will construct the necessary space of cluster functions while ignoring all multi-zeta-values so that we are effectively working at the symbol level. Any beyond-symbol-ambiguities can be easily fixed in the end. 

\subsubsection{Construction of the cluster function space}
The construction of integrable symbols can be done recursively: at weight $k$, we consider all integrable symbols of weight $(k{-}1)$ tensored with elements in the alphabet, and impose integrability conditions on the final two entries. The ansatz reads
$$\sum_{i,j} c_{ij} S^{(k{-}1)}_i \otimes l_j,
$$
where $l_{j=1,\cdots, n_1}$ denote the letters ($n_1$ is the number of letters, or dimension of the cluster algebra), and $S_{i=1,\cdots,N_{k-1}}^{(k{-}1)}$ denote the independent integrable symbols of weight $(k{-}1)$. The integrability condition for the last two entries reads
\begin{equation}\label{integrability}
\sum_{i,j,m} c_{ij} S^{(k{-}2)}_{i;m} d\log l_m \wedge d\log l_j=0,\quad\text{where }S^{(k{-}1)}_i=\sum_mS^{(k{-}2)}_{i;m}\otimes l_m.
\end{equation}
where $S^{(k{-}2)}_{i;m}$ are linear combinations of weight-$(k{-}2)$ integrable symbols.

Before we describe the $A_1^2$ and $A_2$ cluster function spaces of interest, we point out a remarkable fact concerning the number $N_k$ of independent weight-$k$ integrable symbols corresponding to a given alphabet. In general, it is an open question how to compute $N_k$ without explicitly solving the above ansatz at each weight. However, for cluster algebra $A_r$ with no constraints on the alphabet $\{z_i,1{+}z_i,z_j{-}z_i | 1\leq i<j\leq r\}$, there exists a nice recursion relation:
\begin{equation}
    \forall k\geq1:\quad N_k=n_1N_{k-1}-n_2N_{k-2}+\cdots+(-)^{r-1}n_rN_{k-r}.
\end{equation}
Here, $n_m$ is the number of independent $d\log$ $m$-forms, $N_0=1$, and we have defined  $N_{k<0}:=0$ in order to apply it to cases with $k<r$. To see this, define the space $W^{(k,m)}=F^{(k-m)}\otimes_{\mathbb Q}\Omega_{\log}^{(m)}$ of $d\log$ $m$-forms with coefficients being equivalence classes of weight-$(k{-}m)$ pure functions under the symbol map. A theorem (see~\cite{brown2009multiple,Brown:2013qva}) states that for such $A_r$ alphabets, the following sequence is exact:
\begin{equation}
    0\to W^{(k,0)}\xrightarrow{\rm d}\cdots\xrightarrow{\rm d}W^{(k,r)}\xrightarrow{\rm d}\underbrace{W^{(k,r+1)}}_{=0}\xrightarrow{\rm d}\cdots\xrightarrow{\rm d}\underbrace{W^{(k,k)}}_{=0}\to0.
\end{equation}
The recursion relation follows by noticing $N_k=\dim F^{(k)}$ and $n_m=\dim\Omega_{\log}^{(m)}$.

In fact, the recursion relation also holds for ``nice'' degenerations of $A_r$ alphabets. By this, we mean that the imposed constraints commute with the anti-derivative operator. For instance, $A_1^2$ is simply $A_2$ with the letter $v{-}w$ forbidden, but the anti-derivative of every object in $W^{(k,m)}$ satisfying this constraint clearly satisfies the same constraint, so that exactness is not spoiled, as long as we only count the $d\log$ forms without $d\log(v{-}w)$ in computing $n_m$.

The above counting works for type-$A$ cases (and degenerations) where no further constraints have been imposed. When we do impose constraints on possible first entries, things become more subtle due to their possible effect on $n_m$. Fortunately for us, we are concerned with algebras of rank $n=2$ and need only first-entries conditions with $k'=2$, so that starting at $k-r\geq k'$ or $k\geq4$, the coefficients $n_m$ are not affected by such changes several ``entries'' ago.\footnote{Note that such counting only applies to this specific alphabet of type $A$ with linear letters, thus for the familiar $A_3$ alphabet of six-point amplitudes with physical first-entry conditions, where the letters are not linear, our result cannot be applied directly.} Checking $k=3$ explicitly, we conclude that our constraints only change $N_1$ and $N_2$, but do not affect the recursion for $k\geq 3$:
\begin{itemize}
    \item For $A_1^2$, the recursion reads $N_k=4N_{k-1}-4N_{k-2}$ with $k\geq3$;
    \item For $A_2$, the recursion reads $N_k=5N_{k-1}-6N_{k-2}$ with $k\geq3$.
\end{itemize}

For $A_1^2$ and $A_2$ spaces where only $v, 1+v, w, 1+w$ are allowed in the first two entries, we have $N_1=4$ and $N_2=12$. Thus we find that $N_k (A_1^2)=(k+1) 2^k$, and $N_k (A_2)=4 \times 3^{k-1}$. The space from which we bootstrap the amplitudes is spanned by all integrable symbols from two overlapping such $A_2$'s, thus the number at weight $k$ is given by
\begin{equation}
2 N_k(A_2)-N_k(A_1^2)=8\times 3^{k-1}-(k+1) \times 2^k\,.
\end{equation}
Indeed we have explicitly checked that there are $4,12,40,136,456,1496, 4808, 15192$ such integrable symbols at weight $k=1, 2, \cdots, 8$. Given the symbols, it is straightforward to integrate them to obtain actual (2dHPL) functions.

In order to present these functions at each weight in a concise way, we note that it suffices to record only the new functions at each weight, {\it i.e.} those that are not products of lower-weight functions. In practice this can be done by projecting out all functions whose symbol contain shuffle products of lower-weight ones, and we list all new functions up to weight $k=6$ in an ancillary notebook file. 

Let us start with weight $2$, where as mentioned, only $2$ new functions (in addition to the $10$ $\log \log$ functions) are needed, which can be chosen to be ${\rm Li}_2(-v)$ and ${\rm Li}_2(-w)$. At weight $3$, we have $12$ new functions in total: $4$ of which do not contain mixing letters, {\it i.e.} they are $A_1^2$ functions, and they can be chosen as ${\rm Li}_3(-v)$, ${\rm Li}_3(1+v), {\rm Li}_3(-w), {\rm Li}_3(1+w)$; $4$ of them contain letter $v-w$, which are 
\ba
&G_{1,0,0} (\frac{v}{w}),\, G_{w,0,-1}(v)+\log \left(1-\frac{v}{w}\right) G_{-1,0}(w)-\log (w) G_{w,-1}(v),\\\nonumber
&G_{w,-1,0}(v)-\log (w+1) G_{w,0}(v)+\log \left(1-\frac{v}{w}\right) G_{0,-1}(w),\\\nonumber &
G_{w,-1,-1}(v)+\frac{1}{2} \log (w+1)^2 \log \left(1-\frac{v}{w}\right)- \log (w+1) G_{w,-1}(v)
\ea
and the remaining four containing $1-v w$ can be obtained by cyclic rotation $v\to 1/w, w\to v$ (in total we have $12$ new functions). We have included terms that are given by products of lower-weight functions to ensure that only non-mixing letters appear in the first two entries.

Now we move to weight $4,5,6$ {\it etc.}, and again we can write down new $A_1^2$ functions and genuine $A_2$ functions with mixing letters separately. For example at weight $4$, there are $6$ $A_1^2$ functions that are new, which can be chosen as ${\rm Li}_4(-v), {\rm Li}_4(1+v), {\rm Li}_4(1+1/v)$ and $(v\leftrightarrow w)$; there are $12$ new functions which contain the letter $v-w$, and another $12$ obtained by cyclic rotation (containing the letter $1-v w$), thus in total we have $6+12+12=30$ new functions. Similarly at weight $5$, we have $12$ new $A_1^2$ functions, $36$ new functions with $v-w$ and $36$ with $1-v w$, thus $12+36+36=84$ new functions in total; at weight $6$ we find $18+98+98=214$ new functions. To recover the entire function space at each weight, one includes products of lower-weight new functions, {\it e.g.} at weight $3$ in addition to the $12$ new functions, we have $4(4+1)(4+2)/3!=20$ $\log^3$ functions and $4\times 2=8$ $\log {\rm Li}_2$ functions, thus recovering the $40$-dimensional space for $k=3$ as mentioned above. Note that one can also trivially include possible constants such as multi-zeta-values and obtain the full space covering beyond-symbol ambiguities. 

We have not attempted to apply cluster adjacency or extended-Steinmann relations: since the two mixing letters never appear in the same term (let alone next to each other), such constraints can only be studied for the two $A_2$'s separately, but no such constraints has become relevant for the octagon functions we studied, at least not in terms of $v,w$ variables.

With the function space at hand, we are able to bootstrap octagon amplitudes. We will first determine the symbol of amplitudes and fix beyond-symbol-ambiguities in the end. We arrange our results according to the level $k+\ell$, as $\bar Q$ equations relate $\ell$-loop N$^k$MHV amplitudes on the same level. For the $k+\ell=2$ case which turns out to contain only logarithmic terms, symmetry and collinear conditions suffice to determine the answer. For $k+\ell=3$, more conditions are needed, such as those imposed by soft-collinear OPE~\cite{Gaiotto:2010fk}. We summarize our results as follows.

\subsubsection{$k+\ell=2$}

\begin{center}
    \begin{tabular}{||c|cc||}
        & \textbf{1-loop NMHV} & \textbf{2-loop MHV} \\
        \hline
        & 24 & 136 \\
        Last entries & & 80 \\
        Symmetry & 10 & 15 \\
        Collinear & 1 & 1
    \end{tabular}
\end{center}

\paragraph{1-loop NMHV}
From~(\ref{R81}), the symbols to be determined are $\mathcal S(f_{8,1}^1(v,w))$ and $\mathcal S(f_{8,1}^2(v,w))$, which satisfy the following symmetry and collinear conditions:
\begin{equation}
    \begin{gathered}
        f_{8,1}^1(v,w)=f_{8,1}^1(w,v),\\
        \lim_{v\to0}f_{8,1}^1(v,w)=0,
    \end{gathered}
    \quad
    \begin{gathered}
        f_{8,1}^2(v,w)=f_{8,1}^2(w,v)=f_{8,1}^2(1/v,w),\\
        \lim_{v\to\infty} (f_{8,1}^2(v,w)-f_{8,1}^1(v,w)-f_{8,1}^1(v,1/w))=1.
    \end{gathered}
\end{equation}
These suffice to fix the 1-loop NMHV amplitude.

\paragraph{2-loop MHV}
The symbol to be determined is $\mathcal S(R_{8,0}^{(2)}(v,w))$, which satisfies the following symmetry and collinear conditions:
\begin{gather}
    R_{8,0}^{(2)}(v,w)=R_{8,0}^{(2)}(w,1/v)=R_{8,0}^{(2)}(1/v,w),\quad 
    \lim_{v\to0}R_{8,0}^{(2)}(v,w)=1.
\end{gather}

Greater efficiency is achieved by restricting the space of integrable symbols using $\bar Q$ equations, according to the refined version of our conjecture. In particular, for the 2-loop MHV amplitude, mixing letters are not allowed in the 3rd and 4th entries. These conditions are enough to fix the 2-loop MHV amplitude.

\subsubsection{$k+\ell=3$}

\begin{center}
    \begin{tabular}{||c|ccc||}
        & \textbf{1-loop N$^2$MHV} & \textbf{2-loop NMHV} & \textbf{3-loop MHV}\\
        \hline
        & 11 & 272 & 1496 \\
      Last entries & & 192 & 672 \\
        Collinear & 3 & 21 & 40 \\
        {\color{gray}Four-mass} & {\color{gray}1} & & \\
        OPE leading & 1 & 7 & 21 \\
        OPE sub-leading & & 1 & 1
    \end{tabular}
\end{center}

\paragraph{1-loop N$^2$MHV}
From~(\ref{R82}), the symbol to be determined is $\mathcal S(F_{8,2}(v,w))$, which by itself does not satisfy any particular constraints. However, from our analysis above we see that all prefactors are of the form $\frac v{v-w}$. Thus, we may construct from $\mathcal S(F_{8,2}(v,w))$ an ansatz\footnote{As mentioned above, there are 12 integrable symbols at weight-2, so the ansatz for $\mathcal S(F_{8,2}(v,w))$ should have 12 free parameters to begin with. However, after constructing $\mathcal S(R_{8,2}^{(1)}(v,w))$, one of these immediately drops out by symmetry, hence the 11 in the table.} $\mathcal S(R_{8,2}^{(1)}(v,w))$:
\begin{gather}
    \mathcal S(R_{8,2}^{(1)}(v,w))=R_{8,2}^{\text{tree}}(v,w)\left(\frac v{v-w}\mathcal S(F_{8,2}(v,w))+\text{cyclic}\right).
\end{gather}

Now, we can impose the symmetry and collinear conditions:
\begin{gather}
    R_{8,2}^{(1)}(v,w)=R_{8,2}^{(1)}(w,1/v)=R_{8,2}^{(1)}(1/v,w),\quad
    \lim_{v\to0} R_{8,2}^{(1)}(v,w)=0.
\end{gather}
%
To proceed, we impose the constraint that the soft-collinear OPE of our ansatz should match known results. Recall that the soft-collinear OPE is essentially a Taylor expansion of the analytic part after factoring out the non-analytic part:
\begin{align*}
    \mathcal F(v,w) &= \overbrace{F_m(v,w)\log^mv}^{\text{leading order of OPE}}+F_{m-1}(v,w)\log^{m-1}v+\cdots+F_0(v,w),&\text{where }F_i\text{ is analytic},\\
    &= \left(\sum_{n=0}^{\infty}f_{mn}(w)v^n\right)\log^mv+\cdots+\left(\sum_{n=0}^{\infty}f_{0n}(w)v^n\right).
\end{align*}
Notice that the prefactors are part of the functions $F_i(v,w)$, which are still analytic at $v=0$.

At this stage, we should construct the full function space including constants and compute collinear-soft OPE expansion from there. However, we find it more convenient to compute discontinuities at the symbol level, and fix constants later. Therefore, we work ``modulo beyond-symbol-ambiguities''--start by computing the symbol of ${\cal S}(F_i)$ for $i=0,1, \cdots, m$; then, uplift symbols to functions $F_i$ and expand to $f_{ij} (w)$ to compare with known results.

What are the ``known results"? For the leading order of OPE in the MHV case up to 3 loops, the result has been computed from first principles~\cite{Gaiotto:2010fk}. For higher $k$, a first-principle calculation is not present, but we take the equivalent step by comparing with known results of the amplitude. For instance, $\mathcal S(R_{8,2}^{(1)}(v,w))$ should have the following leading ($m=1$) OPE behavior~\cite{Caron-Huot:2013vda}:
\begin{gather}
    \mathcal S(f_{10}(w))=0,\\
    \mathcal S(f_{1n}(w))=\frac{(-1)^n}n+(-1)^n\sum_{i=1}^{n-1}\frac{(-w)^i+(-w)^{-i}}{n-i}+\frac1{w^n}[1+w]+ w^n\left[\frac{1+w}w\right],
\end{gather}
where $[\cdots]$ denotes the tensor produced by the symbol map, so that $[1+w]$ in $\mathcal S(f_{1n}(w))$ simply means $\log(1+w)$ in $f_{1n}(w)$. This enables us to fix $\mathcal S(R_{8,2}^{(1)}(v,w))$.

As a side note, we can impose a different constraint for the last step. From the box expansion of 1-loop amplitudes, it is clear that the residue at the pole $(1-vw)$ must be proportional to the corresponding four-mass box:
\begin{gather}
    \lim_{v\to1/w}(1-vw)R_{8,2}^{(1)}(v,w)\propto F(1,3,5,7).
\end{gather}
This also enables us to fix $\mathcal S(R_{8,2}^{(1)}(v,w))$. 

\paragraph{2-loop NMHV}
The symbols to be determined are again $\mathcal S(f_{8,1}^1)$ and $\mathcal S(f_{8,1}^2)$, which satisfy the symmetry and collinear conditions as in the 1-loop case. According to our conjecture, $\bar Q$ equations exclude mixing letters from the last entry. Using these conditions alone, the space of integrable symbols is reduced to 21 dimensions.

The leading ($m=1$) OPE behavior should match the known results~\cite{Caron-Huot:2013vda}:
\begin{gather}
    \mathcal S(f_{10}(w))=0,\\
    \mathcal S(f_{1n}(w))=\frac{2(-1)^n}{n}[(1{+}w)\otimes(1{+}w)]-\frac{(-1)^n}{n^2}[1{+}w]+\frac{w^n}{n^2}\biggl[\frac{w}{1{+}w}\biggr]-\frac{(-1)^n}{n^2}\sum_{i=1}^{n-1}\frac{(-w)^i}{n{-}i}.
\end{gather}
This further reduces the space to 7 dimensions.

Next, it is natural to use the sub-leading order of OPE, which has not been computed from first principles. Luckily for us, the 2-loop NMHV amplitude itself is known exactly, ready for direct comparison. This way, we are able to test the constraining power of the sub-leading order of OPE without computing it from first principles. The result is satisfying: the 2-loop NMHV amplitude is complete determined by these conditions.

\paragraph{3-loop MHV}
The symbol to be determined is again $\mathcal S(R_{8,0}^{(2)}(v,w))$, which satisfies the symmetry and collinear conditions as in the 2-loop case. According to our conjecture, mixing letters are not allowed on the last two symbol entries. These constraints reduce the dimension of the space of integrable symbols to 40.

In exactly the same way as the 2-loop NMHV case, imposing leading order OPE constraints reduces the dimension to 21, and imposing sub-leading order OPE constraints finishes the job!

\section{Towards $2 n$-gons: $A_2$ functions and beyond}

Having discussed octagon amplitudes and integrals, it is natural to wonder what we can say about amplitudes of higher points, for $2 n= 10, 12, \cdots$. As mentioned earlier, a remarkable property of \Roneone kinematics is that we can apply the {\it collinear-soft uplift} to obtain higher-point amplitudes from lower objects, which are to be viewed as a function of off-shell points rather than polygons~\cite{Goddard:2012cx,Caron-Huot:2013vda}. The uplifting begins with a function $S_8$ which depends on four (off-shell) points, but may involve higher partial amplitudes like $S_{10},S_{12}$ {\it etc.}, depending on the loop order:
\be R_n=\sum_{a \lhd b \lhd c \lhd d} (-)^{a{+}b{+}c{+}d} S_8(x_a, x_b, x_c, x_d)+\mbox{contributions from $S_{10}, S_{12}$ etc}.\label{generaluplift}\ee
In the first term we sum over all four-mass boxes with labels $1\leq a,b,c,d\leq n$, where $a\lhd b$ means that the indices should be separated by at least 2, $a\leq b-2$ (it is understood that $d$ and $a$ must be separated also, e.g. when $a=1$ we must have $d \leq n-1$).  For example, for $2 n=8$ we have two such $S_8$ contributions
\be
 R_{8} = S_{8}(x_2,x_4,x_6,x_8)+S_{8}(x_1,x_3,x_5,x_7) \,, \label{matching8point}
\ee
and for $2n=10, 12,\cdots$, we have $25, 105, \cdots $ $S_8$ contributions, one for each four-mass box. Remarkably for $k+\ell=2$, the $S_8$ contributions are all one can have, {\it i.e.} $S_{10}=S_{12}=\cdots=0$; for $k+\ell=3$, we find non-vanishing remainders up to $S_{12}$, but very nicely they consist of simple logarithmic and dilogarithmic functions without any mixing letters, for two-loop NMHV and three-loop MHV, respectively~\cite{Caron-Huot:2013vda}. Thus we see that, for $k+\ell \leq 3$, the amplitude is given by a sum of $S_8$ functions labelled by four-mass boxes, plus simpler remainders without any mixing letters!

This result for $k+\ell \leq 3$ immediately implies that up to three-loop MHV and two-loop NMHV, $2n$-gon in \Roneone kinematics consists of a collection of $A_2$ functions, one for each four-mass box (or $S_8$), and a $A_{n{-}3}^2$ function for the remainders which contains no mixing letters. Note that for $2n$ points, there are exactly $N_{4m} (n) \equiv n(n{-}3)(2n{-}5)(2n{-}7)/6$ four-mass boxes (or $S_8$'s), {\it e.g.} $N_{4m}=2, 25, 105, \cdots$ for $n=4,5,6,\cdots$. 
For example, for $2n=10$, there are 25 $A_2$ functions and a $A_2 \times A_2$ function (note different meanings of these $A_2$'s), thus 35 letters in total; for $2n=12$, we have $123$ letters, or 105 $A_2$ functions and a $A_3^2$  function. Purely for the alphabet, the $N_{4m}(n)$ $A_2$'s already include every letter in the $A_{n{-}3}^2$ part, but we emphasize that for $2n \geq 10$, in addition to the $A_2$ functions (one for each $S_8$), we still need an $A_{n{-}3}^2$ function for the remainder which cannot be split into various $A_2$. 

In general it would be more difficult to bootstrap for $2n \geq 10$ with this alphabet since the space of integrable symbol grows faster. Let us briefly comment on one loop (weight $2$) case. Note that only the $A_{n{-}3}^2$ part can appear for one loop, and it is straightforward to enumerate weight-2 integrable symbols (or functions). There are two possibilities: ${\rm Li}_2$ functions and $\log \log$ ones; the former are single-variable functions which can be listed in even and odd sectors separately, and the latter are given by the symmetric part ${\cal S}[\log a \log b]=a\otimes b + b \otimes a$ with $a, b \in A_{n{-}3}^2$. Note that the point-count of the cluster configuration space of $A_{n{-}3}$ on $\mathbb{F}_p$ is given by a polynomial $(p-2)(p-3) \cdots (p-n+2)$ for a generic prime number $p$, and we can infer the number of independent ${\rm d}\log$ $k$-forms from the coefficient of $p^{n{-}3{-}k}$ (up to a sign); In particular, the number of $1$ forms is $n_1:=n(n{-}3)/2$ and the number of $2$ forms is $n_2=(3n^2-n+2)(n-3)(n-4)/24$. 

It is clear that we have $n_1(2 n_1+1)$ symmetric symbols (or $\log \log$ functions), and the enumeration of ${\rm Li}_2$ functions in each sector is as follows:
from the $n_1$ letters of $A_{n-3}$, we can construct ${n_1 \choose 2}$ antisymmetric weight-$2$ symbols; on the other hand, integrability exactly imposes $n_2$ constraints; thus there are ${n_1\choose2}-n_2={n-1\choose3}$ antisymmetric integrable weight-2 symbols, or ${\rm Li}_2$ functions\footnote{It is interesting that this is also the number of independent \Roneone four-mass box functions in each sector, as one can shown by repeatedly using Abel identities for Rogers dilogarithms.}. For example, for $2n=8$ we have $1+1$ ${\rm Li}_2$ functions and $2\times 5$ $\log \log$'s, for $2n=10$, we have $4+4$ ${\rm Li}_2$'s and $5\times 11$ $\log \log$'s, and for $2n=12$, we have $10+10$ ${\rm Li}_2$'s and $9\times 19$ $\log \log$'s. 

\begin{figure}[htbp]
\centering
\subfigure[$I^2_{\rm dp}(1,5,5,8,10)$]{
\begin{tikzpicture}[baseline={([yshift=-.5ex]current bounding box.center)},scale=0.25]
\draw[black,thick](0,0)--(0,5)--(4.75,6.54)--(7.69,2.50)--(4.75,-1.54)--cycle;
\draw[black,thick](0,5)--(-4.75,6.54)--(-7.69,2.50)--(-4.75,-1.54)--(0,0);
\draw[black,thick](4.75,6.54)--(6,7.94);
\draw[black,thick](4.75,-1.54)--(6,-3.04);
\draw[black,thick](9.19,1.6)--(7.69,2.50)--(9.19,3.4);
\draw[black,thick](9.19,2.5)--(7.69,2.50);
\draw[black,thick] (-4.65,8.50)--(-4.75,6.54)--(-6,7.94);
\draw[black,thick] (-4.65,-3.50)--(-4.75,-1.54)--(-6,-3.04);
\draw[black,thick](-9.19,2.5)--(-7.69,2.50);
\filldraw[black] (6,7.94) node[anchor=south west] {{$1$}};
\filldraw[black] (9.19,3.4) node[anchor=west] {{$2$}};
\filldraw[black] (9.19,2.5) node[anchor=west] {{$3$}};
\filldraw[black] (9.19,1.6) node[anchor=west] {{$4$}};
\filldraw[black] (6,-3.04) node[anchor=north west] {{$5$}};
\filldraw[black] (-5,8.50) node[anchor=south west] {{$10$}};
\filldraw[black] (-6,7.94) node[anchor=south east] {{$9$}};
\filldraw[black] (-9.19,2.5) node[anchor=east] {{$8$}};
\filldraw[black] (-5,-3.50) node[anchor=north west] {{$6$}};
\filldraw[black] (-6,-3.04) node[anchor=north east] {{$7$}};
\filldraw[black] (3.5,1.04) node[anchor=south] {{${\bf N}_1$}};
\filldraw[black] (-3.5,1.04) node[anchor=south] {{${\bf N}_2$}};
\end{tikzpicture}
}
\subfigure[$I^2_{\rm dp}(1,4,4,7,10)$]{
\begin{tikzpicture}[baseline={([yshift=-.5ex]current bounding box.center)},scale=0.25]
\draw[black,thick](0,0)--(0,5)--(4.75,6.54)--(7.69,2.50)--(4.75,-1.54)--cycle;
\draw[black,thick](0,5)--(-4.75,6.54)--(-7.69,2.50)--(-4.75,-1.54)--(0,0);
\draw[black,thick](4.75,6.54)--(6,7.94);
\draw[black,thick](4.75,-1.54)--(6,-3.04);
\draw[black,thick](9.19,1.6)--(7.69,2.50)--(9.19,3.4);
\draw[black,thick] (-4.65,8.50)--(-4.75,6.54)--(-6,7.94);
\draw[black,thick] (-3.7,8.20)--(-4.75,6.54);
\draw[black,thick] (-4.65,-3.50)--(-4.75,-1.54)--(-6,-3.04);
\draw[black,thick](-9.19,2.5)--(-7.69,2.50);
\filldraw[black] (6,7.94) node[anchor=south west] {{$1$}};
\filldraw[black] (9.19,3.4) node[anchor=west] {{$2$}};
\filldraw[black] (9.19,1.6) node[anchor=west] {{$3$}};
\filldraw[black] (6,-3.04) node[anchor=north west] {{$4$}};
\filldraw[black] (-5,8.50) node[anchor=south] {{$9$}};
\filldraw[black] (-3.7,8.20) node[anchor=south west] {{$10$}};
\filldraw[black] (-6,7.94) node[anchor=south east] {{$8$}};
\filldraw[black] (-9.19,2.5) node[anchor=east] {{$7$}};
\filldraw[black] (-5,-3.50) node[anchor=north west] {{$5$}};
\filldraw[black] (-6,-3.04) node[anchor=north east] {{$6$}};
\filldraw[black] (3.5,1.04) node[anchor=south] {{${\bf N}_1$}};
\filldraw[black] (-3.5,1.04) node[anchor=south] {{${\bf N}_2$}};
\end{tikzpicture}
}
\caption{Two examples of 10pt integrals}\label{fig:10pt}
\end{figure}
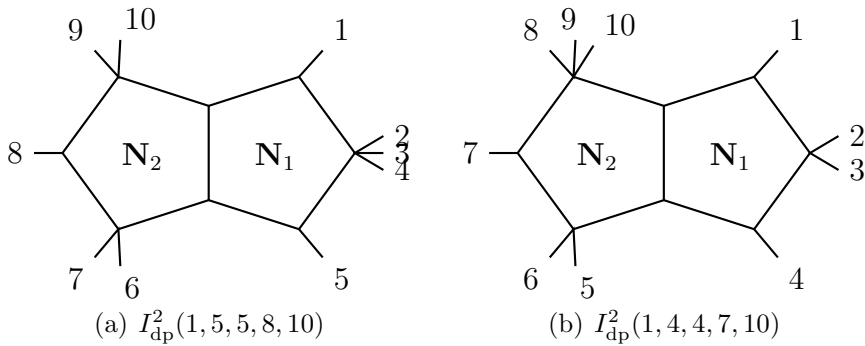

Besides amplitudes with $k+\ell \leq 3$, we find that simple Feynman integrals such as double pentagons with $2n \geq 10$ legs have similar alphabets in \Roneone kinematics. We can see this already for integrals without four-mass square roots. The simplest $2n=10$ point cases are the following two double-pentagon integrals (Fig.~\ref{fig:10pt}). Following the configuration of external legs, we see the left one depends on three cross ratios $\{v=\frac{\langle57\rangle\langle19\rangle}{\langle59\rangle\langle17\rangle},w_1=\frac{[26][8\,10]}{[68][2\,10]},w_2=\frac{[4\,10][68]}{[8\,10][46]}\}$, while the right one $\{v=\frac{\langle17\rangle\langle35\rangle}{\langle13\rangle\langle57\rangle}, w_1=\frac{[26][4\,10]}{[24][6\,10]}, w_2=\frac{[28][4\,10]}{[24][8\,10]}\}$.  Explicit computation shows that the alphabet of the integrals are both $\{v,1+v,w_1,w_2,1+w_1,1+w_2,1-w_1 w_2\}$. Hence both integrals are $A_2\otimes A_1$ functions. For integrals with square roots, we find that in addition to such $A_{n{-}3}^2$ part, the part with algebraic letters again reduces to $A_2$ symbols with mixing letters, one for each square root. For example, for the most general double pentagon with $2n=12$ points, its algebraic words become the sum of $16$ $A_2$ symbols, of the form ${\cal S}(F(v,w))\otimes (v-w) \otimes \frac{v}{w}$, where $v,w$ are the \Roneone cross-ratios of the corresponding four-mass box configuration. 

What do we expect for higher $k$ and/or higher loops? It is well known that even at two loops, amplitudes in 4d with $k\geq 3$ involve functions that go beyond multiple polylogarithms (MPL)~\cite{CaronHuot:2012ab,Bourjaily:2017bsb}, but we do not know if this is the case in \Roneone kinematics (we have not excluded the possibility that \Roneone amplitudes still evaluate to MPL for $k\geq 3$). For $k\leq 2$, it is plausible that amplitudes evaluate to MPLs only; in these cases if one can prove that in 4d, all algebraic letters for amplitudes with $k\leq 2$ are still associated with square roots of four-mass boxes, then the above conjecture holds as well. Starting $k=3$, it is likely that more complicated algebraic singularities in 4d lead to new letters in \Roneone kinematics. In other words, for $k\geq 3$ it is possible that remainder functions $S_{10}, S_{12}$ {\it etc.} contain new mixing letters associated with more legs (thus depending on more than just a pair of $v$ and $w$), which represent more general algebraic letters in 4d. It would be extremely interesting to find amplitudes with such mixing letters, and study what kind of cluster functions, beyond $A_2$ and $A_{n{-}3}^2$, are needed for these new beasts.

\section{Conclusion and discussions}

In this paper, we have studied scattering amplitudes and Feynman integrals in ${\cal N}=4$ SYM in \Roneone kinematics by exploring and leveraging their remarkably simple cluster-algebra structures. For the octagon, we have used the alphabet of two-overlapping $A_2$'s and conditions on first and last entries to bootstrap amplitudes up to $k+\ell=3$. We have also studied various octagon integrals, which are all $A_2$ functions with interesting patterns of the symbol regarding such mixing letters; these allow us to predict patterns for algebraic letters in 4d. We have also discussed generalization to $2n$-gon using $A_2$ functions and cluster functions involving more legs.

There are several natural directions for future investigations. Among others, it would be highly desirable to determine what functions appear for \Roneone amplitudes with $2n\geq 10$ and $k\geq 3$. Some components of 4d super-amplitudes/WL, which are given by elliptic integrals and beyond {\it e.g.} ``train tracks" in~\cite{Bourjaily:2018ycu}, may vanish when supersymmetrically reduced to \Roneone. For example, the famous ``elliptic" component of two-loop N$^3$MHV amplitude~\cite{CaronHuot:2012ab} vanishes since there are not enough $\eta$'s for it in supersymmetric \Roneone. It would be interesting to see if similar components which are given by functions beyond MPL survive the supersymmetric \Roneone reduction. Relatedly, it is natural to expect that algebraic singularities beyond four-mass-box square roots still reduce to rational, mixing letters in \Roneone kinematics for $2n\geq 10$, and it would be interesting to study what type of cluster algebras they belong to.

We have only scratched the surface of the analytic formulas for amplitudes and integrals in \Roneone kinematics. Although in principle they can be obtained by reduction from 4d, the limit is very subtle and it is preferable to study them directly in 2d. This is already the case for leading singularities: the corresponding positroids in 4d usually do not have a well-defined 2d limit, which makes the reduction quite subtle even at one loop! We would like to develop a systematic way for classifying all Yangian invariants in \Roneone kinematics: they must be given by products of even and odd $m=2$ R-invariants, times possible prefactors that mix them. Based on this, one can study amplitudes/WL for $k+\ell \geq 3$ and Feynman integrals at higher loops, where \Roneone kinematics serves as a perfect laboratory. We can hope to even address questions such as resummation of all-loop integrals/amplitudes, and make connection with strong coupling results~\cite{Alday:2009ga,Alday:2009yn} as discussed in~\cite{Caron-Huot:2013vda}. 

For bootstrapping higher-loop octagons, we remark that OPE has not been properly understood so far in \Roneone kinematics (again the reduction to 2d is very subtle in the expansion), and it would be highly desirable to study this more systematically. Among other goals, combined with constraints from Yangian equations {\it etc.}, it is possible to push the frontier of octagon bootstrap further. A very concrete target is the bootstrap of two-loop N$^2$MHV octagon, where $\bar Q$ equations alone do not provide any additional constraints, but it is plausible that the first few orders of OPE data is sufficient to determine the amplitude. We can then go to higher points using collinear-soft uplift, and again by  $\bar Q$ that would allow us to reach the land of $k+\ell=4$ amplitudes, including three-loop NMHV and four-loop MHV cases. 

Last but not least, there is the interesting question of constructing the space of integrable symbols and determining the dimension at each weight, for a given alphabet and possible constraints. We have only studied the simplest cases with {\it linear} alphabets of type $A$, where if we do not impose any other constraints, $N_k$ can be determined purely from $n_m$ (the number of independent $d\log$ $m$-forms) for $m\leq k$. It is already interesting to see 
that the recursion still works if we have imposed certain constraints for first $k'$ entries. We suspect that exactly the same argument can be applied to any alphabet with linear letters (corresponding to hyperplane arrangements), and similar recursion relations should hold for those cases. We leave the question of extending these results to other alphabets (e.g., other finite-type cluster algebras) to future investigations. 

\acknowledgments

It is a pleasure to thank Nima Arkani-Hamed, Livia Ferro, Johannes Henn, Yu-tin Huang, Chia-Kai Kuo and Chi Zhang for encouraging discussions, correspondence and collaborations on related projects. This research is supported in part by National Natural Science Foundation of China under Grant No. 11935013, 11947301, 12047502, 12047503.

\bibliographystyle{utphys}
\bibliography{main}

\providecommand{\noopsort}[1]{}\providecommand{\singleletter}[1]{#1}%
\providecommand{\href}[2]{#2}\begingroup\raggedright\begin{thebibliography}{10}

\bibitem{ArkaniHamed:2010kv}
N.~Arkani-Hamed, J.~L. Bourjaily, F.~Cachazo, S.~Caron-Huot, and J.~Trnka,
  ``{The All-Loop Integrand For Scattering Amplitudes in Planar N=4 SYM},''
  \href{http://dx.doi.org/10.1007/JHEP01(2011)041}{{\em JHEP} {\bfseries 01}
  (2011) 041},
\href{http://arxiv.org/abs/1008.2958}{{\ttfamily arXiv:1008.2958 [hep-th]}}.

\bibitem{Arkani-Hamed:2016byb}
N.~Arkani-Hamed, J.~L. Bourjaily, F.~Cachazo, A.~B. Goncharov, A.~Postnikov,
  and J.~Trnka, \href{http://dx.doi.org/10.1017/CBO9781316091548}{{\em
  {Grassmannian Geometry of Scattering Amplitudes}}}.
\newblock Cambridge University Press, 4, 2016.
\newblock \href{http://arxiv.org/abs/1212.5605}{{\ttfamily arXiv:1212.5605
  [hep-th]}}.

\bibitem{Arkani-Hamed:2013jha}
N.~Arkani-Hamed and J.~Trnka, ``{The Amplituhedron},''
  \href{http://dx.doi.org/10.1007/JHEP10(2014)030}{{\em JHEP} {\bfseries 10}
  (2014) 030},
\href{http://arxiv.org/abs/1312.2007}{{\ttfamily arXiv:1312.2007 [hep-th]}}.

\bibitem{Dixon:2011pw}
L.~J. Dixon, J.~M. Drummond, and J.~M. Henn, ``{Bootstrapping the three-loop
  hexagon},'' \href{http://dx.doi.org/10.1007/JHEP11(2011)023}{{\em JHEP}
  (2011) 023},
\href{http://arxiv.org/abs/1108.4461}{{\ttfamily arXiv:1108.4461 [hep-th]}}.

\bibitem{Dixon:2014xca}
L.~J. Dixon, J.~M. Drummond, C.~Duhr, M.~von Hippel, and J.~Pennington,
  ``{Bootstrapping six-gluon scattering in planar N=4 super-Yang-Mills
  theory},'' \href{http://dx.doi.org/10.22323/1.211.0077}{{\em PoS} {\bfseries
  LL2014} (2014) 077},
\href{http://arxiv.org/abs/1407.4724}{{\ttfamily arXiv:1407.4724 [hep-th]}}.

\bibitem{Dixon:2014iba}
L.~J. Dixon and M.~von Hippel, ``{Bootstrapping an NMHV amplitude through three
  loops},'' \href{http://dx.doi.org/10.1007/JHEP10(2014)065}{{\em JHEP}
  {\bfseries 10} (2014) 065},
\href{http://arxiv.org/abs/1408.1505}{{\ttfamily arXiv:1408.1505 [hep-th]}}.

\bibitem{Drummond:2014ffa}
J.~M. Drummond, G.~Papathanasiou, and M.~Spradlin, ``{A Symbol of Uniqueness:
  The Cluster Bootstrap for the 3-Loop MHV Heptagon},''
  \href{http://dx.doi.org/10.1007/JHEP03(2015)072}{{\em JHEP} {\bfseries 03}
  (2015) 072},
\href{http://arxiv.org/abs/1412.3763}{{\ttfamily arXiv:1412.3763 [hep-th]}}.

\bibitem{Dixon:2015iva}
L.~J. Dixon, M.~von Hippel, and A.~J. McLeod, ``{The four-loop six-gluon NMHV
  ratio function},'' \href{http://dx.doi.org/10.1007/JHEP01(2016)053}{{\em
  JHEP} {\bfseries 01} (2016) 053},
\href{http://arxiv.org/abs/1509.08127}{{\ttfamily arXiv:1509.08127 [hep-th]}}.

\bibitem{Caron-Huot:2016owq}
S.~Caron-Huot, L.~J. Dixon, A.~McLeod, and M.~von Hippel, ``{Bootstrapping a
  Five-Loop Amplitude Using Steinmann Relations},''
  \href{http://dx.doi.org/10.1103/PhysRevLett.117.241601}{{\em Phys. Rev.
  Lett.} {\bfseries 117} no.~24, (2016) 241601},
\href{http://arxiv.org/abs/1609.00669}{{\ttfamily arXiv:1609.00669 [hep-th]}}.

\bibitem{Dixon:2016nkn}
L.~J. Dixon, J.~Drummond, T.~Harrington, A.~J. McLeod, G.~Papathanasiou, and
  M.~Spradlin, ``{Heptagons from the Steinmann Cluster Bootstrap},''
  \href{http://dx.doi.org/10.1007/JHEP02(2017)137}{{\em JHEP} {\bfseries 02}
  (2017) 137},
\href{http://arxiv.org/abs/1612.08976}{{\ttfamily arXiv:1612.08976 [hep-th]}}.

\bibitem{Drummond:2018caf}
J.~Drummond, J.~Foster, {\"{O}}.~G{\"{u}}rdo{\u{g}}an, and G.~Papathanasiou,
  ``{Cluster adjacency and the four-loop NMHV heptagon},''
  \href{http://dx.doi.org/10.1007/JHEP03(2019)087}{{\em JHEP} {\bfseries 03}
  (2019) 087},
\href{http://arxiv.org/abs/1812.04640}{{\ttfamily arXiv:1812.04640 [hep-th]}}.

\bibitem{Caron-Huot:2019vjl}
S.~Caron-Huot, L.~J. Dixon, F.~Dulat, M.~von Hippel, A.~J. McLeod, and
  G.~Papathanasiou, ``{Six-Gluon amplitudes in planar $ \mathcal{N} $ = 4
  super-Yang-Mills theory at six and seven loops},''
  \href{http://dx.doi.org/10.1007/JHEP08(2019)016}{{\em JHEP} {\bfseries 08}
  (2019) 016},
\href{http://arxiv.org/abs/1903.10890}{{\ttfamily arXiv:1903.10890 [hep-th]}}.

\bibitem{Caron-Huot:2019bsq}
S.~Caron-Huot, L.~J. Dixon, F.~Dulat, M.~von Hippel, A.~J. McLeod, and
  G.~Papathanasiou, ``{The Cosmic Galois Group and Extended Steinmann Relations
  for Planar $\mathcal{N} = 4$ SYM Amplitudes},''
  \href{http://dx.doi.org/10.1007/JHEP09(2019)061}{{\em JHEP} {\bfseries 09}
  (2019) 061},
\href{http://arxiv.org/abs/1906.07116}{{\ttfamily arXiv:1906.07116 [hep-th]}}.

\bibitem{Dixon:2020cnr}
L.~J. Dixon and Y.-T. Liu, ``{Lifting Heptagon Symbols to Functions},''
  \href{http://dx.doi.org/10.1007/JHEP10(2020)031}{{\em JHEP} {\bfseries 10}
  (2020) 031}, \href{http://arxiv.org/abs/2007.12966}{{\ttfamily
  arXiv:2007.12966 [hep-th]}}.

\bibitem{Chicherin:2017dob}
D.~Chicherin, J.~Henn, and V.~Mitev, ``{Bootstrapping pentagon functions},''
  \href{http://dx.doi.org/10.1007/JHEP05(2018)164}{{\em JHEP} {\bfseries 05}
  (2018) 164}, \href{http://arxiv.org/abs/1712.09610}{{\ttfamily
  arXiv:1712.09610 [hep-th]}}.

\bibitem{Caron-Huot:2020bkp}
S.~Caron-Huot, L.~J. Dixon, J.~M. Drummond, F.~Dulat, J.~Foster,
  O.~G\"urdo\u{g}an, M.~von Hippel, A.~J. McLeod, and G.~Papathanasiou, ``{The
  Steinmann Cluster Bootstrap for $N$ = 4 Super Yang-Mills Amplitudes},''
  \href{http://dx.doi.org/10.22323/1.376.0003}{{\em PoS} {\bfseries CORFU2019}
  (2020) 003}, \href{http://arxiv.org/abs/2005.06735}{{\ttfamily
  arXiv:2005.06735 [hep-th]}}.

\bibitem{Golden:2013xva}
J.~Golden, A.~B. Goncharov, M.~Spradlin, C.~Vergu, and A.~Volovich, ``{Motivic
  Amplitudes and Cluster Coordinates},''
  \href{http://dx.doi.org/10.1007/JHEP01(2014)091}{{\em JHEP} {\bfseries 01}
  (2014) 091},
\href{http://arxiv.org/abs/1305.1617}{{\ttfamily arXiv:1305.1617 [hep-th]}}.

\bibitem{Goncharov:2010jf}
A.~B. Goncharov, M.~Spradlin, C.~Vergu, and A.~Volovich, ``{Classical
  Polylogarithms for Amplitudes and Wilson Loops},''
  \href{http://dx.doi.org/10.1103/PhysRevLett.105.151605}{{\em Phys. Rev.
  Lett.} {\bfseries 105} (2010) 151605},
\href{http://arxiv.org/abs/1006.5703}{{\ttfamily arXiv:1006.5703 [hep-th]}}.

\bibitem{Duhr:2011zq}
C.~Duhr, H.~Gangl, and J.~R. Rhodes, ``{From polygons and symbols to
  polylogarithmic functions},''
  \href{http://dx.doi.org/10.1007/JHEP10(2012)075}{{\em JHEP} {\bfseries 10}
  (2012) 075},
\href{http://arxiv.org/abs/1110.0458}{{\ttfamily arXiv:1110.0458 [math-ph]}}.

\bibitem{fomin2002cluster}
S.~Fomin and A.~Zelevinsky, ``Cluster algebras i: foundations,'' {\em Journal
  of the American Mathematical Society} {\bfseries 15} no.~2, (2002) 497--529.

\bibitem{fomin2003cluster}
S.~Fomin and A.~Zelevinsky, ``Cluster algebras ii: Finite type
  classification,'' {\em Inventiones mathematicae} {\bfseries 154} no.~1,
  (2003) 63--121.

\bibitem{speyer2005tropical}
D.~Speyer and L.~Williams, ``The tropical totally positive grassmannian,'' {\em
  Journal of Algebraic Combinatorics} {\bfseries 22} no.~2, (2005) 189--210.

\bibitem{newpaper}
Z.~Li and C.~Zhang to appear.

\bibitem{Zhang:2019vnm}
S.~He, Z.~Li, and C.~Zhang, ``{Two-loop Octagons, Algebraic Letters and
  $\bar{Q}$ Equations},''
  \href{http://dx.doi.org/10.1103/PhysRevD.101.061701}{{\em Phys. Rev. D}
  {\bfseries 101} no.~6, (2020) 061701},
  \href{http://arxiv.org/abs/1911.01290}{{\ttfamily arXiv:1911.01290
  [hep-th]}}.

\bibitem{He:2020vob}
S.~He, Z.~Li, and C.~Zhang, ``{The symbol and alphabet of two-loop NMHV
  amplitudes from $\bar{Q}$ equations},''
  \href{http://arxiv.org/abs/2009.11471}{{\ttfamily arXiv:2009.11471
  [hep-th]}}.

\bibitem{CaronHuot:2011kk}
S.~Caron-Huot and S.~He, ``{Jumpstarting the All-Loop S-Matrix of Planar N=4
  Super Yang-Mills},'' \href{http://dx.doi.org/10.1007/JHEP07(2012)174}{{\em
  JHEP} {\bfseries 07} (2012) 174},
\href{http://arxiv.org/abs/1112.1060}{{\ttfamily arXiv:1112.1060 [hep-th]}}.

\bibitem{Drummond:2019cxm}
J.~Drummond, J.~Foster, O.~G\"urdogan, and C.~Kalousios, ``{Algebraic
  singularities of scattering amplitudes from tropical geometry},''
  \href{http://arxiv.org/abs/1912.08217}{{\ttfamily arXiv:1912.08217
  [hep-th]}}.

\bibitem{Henke:2019hve}
N.~Henke and G.~Papathanasiou, ``{How tropical are seven- and eight-particle
  amplitudes?},'' \href{http://dx.doi.org/10.1007/JHEP08(2020)005}{{\em JHEP}
  {\bfseries 08} (2020) 005}, \href{http://arxiv.org/abs/1912.08254}{{\ttfamily
  arXiv:1912.08254 [hep-th]}}.

\bibitem{Arkani-Hamed:2019rds}
N.~Arkani-Hamed, T.~Lam, and M.~Spradlin, ``{Non-perturbative geometries for
  planar $\mathcal{N}=4$ SYM amplitudes},''
  \href{http://arxiv.org/abs/1912.08222}{{\ttfamily arXiv:1912.08222
  [hep-th]}}.

\bibitem{Herderschee:2021dez}
A.~Herderschee, ``{Algebraic branch points at all loop orders from positive
  kinematics and wall crossing},''
  \href{http://arxiv.org/abs/2102.03611}{{\ttfamily arXiv:2102.03611
  [hep-th]}}.

\bibitem{Mago:2020kmp}
J.~Mago, A.~Schreiber, M.~Spradlin, and A.~Volovich, ``{Symbol alphabets from
  plabic graphs},'' \href{http://dx.doi.org/10.1007/JHEP10(2020)128}{{\em JHEP}
  {\bfseries 10} (2020) 128}, \href{http://arxiv.org/abs/2007.00646}{{\ttfamily
  arXiv:2007.00646 [hep-th]}}.

\bibitem{He:2020uhb}
S.~He and Z.~Li, ``{A Note on Letters of Yangian Invariants},''
  \href{http://dx.doi.org/10.1007/JHEP02(2021)155}{{\em JHEP} {\bfseries 02}
  (2021) 155}, \href{http://arxiv.org/abs/2007.01574}{{\ttfamily
  arXiv:2007.01574 [hep-th]}}.

\bibitem{Mago:2020nuv}
J.~Mago, A.~Schreiber, M.~Spradlin, A.~Yelleshpur~Srikant, and A.~Volovich,
  ``{Symbol Alphabets from Plabic Graphs II: Rational Letters},''
  \href{http://arxiv.org/abs/2012.15812}{{\ttfamily arXiv:2012.15812
  [hep-th]}}.

\bibitem{Caron-Huot:2018dsv}
S.~Caron-Huot, L.~J. Dixon, M.~von Hippel, A.~J. McLeod, and G.~Papathanasiou,
  ``{The Double Pentaladder Integral to All Orders},''
  \href{http://dx.doi.org/10.1007/JHEP07(2018)170}{{\em JHEP} {\bfseries 07}
  (2018) 170},
\href{http://arxiv.org/abs/1806.01361}{{\ttfamily arXiv:1806.01361 [hep-th]}}.

\bibitem{He:2021esx}
S.~He, Z.~Li, and Q.~Yang, ``{Notes on cluster algebras and some all-loop
  Feynman integrals},'' \href{http://arxiv.org/abs/2103.02796}{{\ttfamily
  arXiv:2103.02796 [hep-th]}}.

\bibitem{Chicherin:2020umh}
D.~Chicherin, J.~M. Henn, and G.~Papathanasiou, ``{Cluster algebras for Feynman
  integrals},'' \href{http://arxiv.org/abs/2012.12285}{{\ttfamily
  arXiv:2012.12285 [hep-th]}}.

\bibitem{He:2020lcu}
S.~He, Z.~Li, Q.~Yang, and C.~Zhang, ``{Feynman Integrals and Scattering
  Amplitudes from Wilson Loops},''
  \href{http://arxiv.org/abs/2012.15042}{{\ttfamily arXiv:2012.15042
  [hep-th]}}.

\bibitem{Drummond:2017ssj}
J.~Drummond, J.~Foster, and {\"{O}}.~G{\"{u}}rdo{\u{g}}an, ``{Cluster Adjacency
  Properties of Scattering Amplitudes in $N=4$ Supersymmetric Yang-Mills
  Theory},'' \href{http://dx.doi.org/10.1103/PhysRevLett.120.161601}{{\em Phys.
  Rev. Lett.} {\bfseries 120} no.~16, (2018) 161601},
\href{http://arxiv.org/abs/1710.10953}{{\ttfamily arXiv:1710.10953 [hep-th]}}.

\bibitem{Alday:2009ga}
L.~F. Alday and J.~Maldacena, ``{Minimal surfaces in AdS and the eight-gluon
  scattering amplitude at strong coupling},''
  \href{http://arxiv.org/abs/0903.4707}{{\ttfamily arXiv:0903.4707 [hep-th]}}.

\bibitem{Alday:2009yn}
L.~F. Alday and J.~Maldacena, ``{Null polygonal Wilson loops and minimal
  surfaces in Anti-de-Sitter space},''
  \href{http://dx.doi.org/10.1088/1126-6708/2009/11/082}{{\em JHEP} {\bfseries
  11} (2009) 082}, \href{http://arxiv.org/abs/0904.0663}{{\ttfamily
  arXiv:0904.0663 [hep-th]}}.

\bibitem{DelDuca:2010zp}
V.~Del~Duca, C.~Duhr, and V.~A. Smirnov, ``{A Two-Loop Octagon Wilson Loop in N
  = 4 SYM},'' \href{http://dx.doi.org/10.1007/JHEP09(2010)015}{{\em JHEP}
  {\bfseries 09} (2010) 015}, \href{http://arxiv.org/abs/1006.4127}{{\ttfamily
  arXiv:1006.4127 [hep-th]}}.

\bibitem{Heslop:2010kq}
P.~Heslop and V.~V. Khoze, ``{Analytic Results for MHV Wilson Loops},''
  \href{http://dx.doi.org/10.1007/JHEP11(2010)035}{{\em JHEP} {\bfseries 11}
  (2010) 035}, \href{http://arxiv.org/abs/1007.1805}{{\ttfamily arXiv:1007.1805
  [hep-th]}}.

\bibitem{Caron-Huot:2013vda}
S.~Caron-Huot and S.~He, ``{Three-loop octagons and $n$-gons in maximally
  supersymmetric Yang-Mills theory},''
  \href{http://dx.doi.org/10.1007/JHEP08(2013)101}{{\em JHEP} {\bfseries 08}
  (2013) 101},
\href{http://arxiv.org/abs/1305.2781}{{\ttfamily arXiv:1305.2781 [hep-th]}}.

\bibitem{Goddard:2012cx}
T.~Goddard, P.~Heslop, and V.~V. Khoze, ``{Uplifting Amplitudes in Special
  Kinematics},'' \href{http://dx.doi.org/10.1007/JHEP10(2012)041}{{\em JHEP}
  {\bfseries 10} (2012) 041}, \href{http://arxiv.org/abs/1205.3448}{{\ttfamily
  arXiv:1205.3448 [hep-th]}}.

\bibitem{Gehrmann:2001jv}
T.~Gehrmann and E.~Remiddi, ``{Numerical evaluation of two-dimensional harmonic
  polylogarithms},''
  \href{http://dx.doi.org/10.1016/S0010-4655(02)00139-X}{{\em Comput. Phys.
  Commun.} {\bfseries 144} (2002) 200--223},
  \href{http://arxiv.org/abs/hep-ph/0111255}{{\ttfamily arXiv:hep-ph/0111255}}.

\bibitem{Torres:2013vba}
M.~A.~C. Torres, ``{Cluster algebras in scattering amplitudes with special 2D
  kinematics},'' \href{http://dx.doi.org/10.1140/epjc/s10052-014-2757-y}{{\em
  Eur. Phys. J. C} {\bfseries 74} (2014) 2757},
  \href{http://arxiv.org/abs/1310.6906}{{\ttfamily arXiv:1310.6906 [hep-th]}}.

\bibitem{He:2020uxy}
S.~He, Z.~Li, Y.~Tang, and Q.~Yang, ``{The Wilson-loop $d \log$ representation
  for Feynman integrals},'' \href{http://arxiv.org/abs/2012.13094}{{\ttfamily
  arXiv:2012.13094 [hep-th]}}.

\bibitem{Ferro:2012wa}
L.~Ferro, ``{Differential equations for multi-loop integrals and
  two-dimensional kinematics},''
  \href{http://dx.doi.org/10.1007/JHEP04(2013)160}{{\em JHEP} {\bfseries 04}
  (2013) 160}, \href{http://arxiv.org/abs/1204.1031}{{\ttfamily arXiv:1204.1031
  [hep-th]}}.

\bibitem{Alday:2010ku}
L.~F. Alday, D.~Gaiotto, J.~Maldacena, A.~Sever, and P.~Vieira, ``{An Operator
  Product Expansion for Polygonal null Wilson Loops},''
  \href{http://dx.doi.org/10.1007/JHEP04(2011)088}{{\em JHEP} {\bfseries 04}
  (2011) 088}, \href{http://arxiv.org/abs/1006.2788}{{\ttfamily arXiv:1006.2788
  [hep-th]}}.

\bibitem{Gaiotto:2010fk}
D.~Gaiotto, J.~Maldacena, A.~Sever, and P.~Vieira, ``{Bootstrapping Null
  Polygon Wilson Loops},''
  \href{http://dx.doi.org/10.1007/JHEP03(2011)092}{{\em JHEP} {\bfseries 03}
  (2011) 092}, \href{http://arxiv.org/abs/1010.5009}{{\ttfamily arXiv:1010.5009
  [hep-th]}}.

\bibitem{Gaiotto:2011dt}
D.~Gaiotto, J.~Maldacena, A.~Sever, and P.~Vieira, ``{Pulling the straps of
  polygons},'' \href{http://dx.doi.org/10.1007/JHEP12(2011)011}{{\em JHEP}
  {\bfseries 12} (2011) 011},
\href{http://arxiv.org/abs/1102.0062}{{\ttfamily arXiv:1102.0062 [hep-th]}}.

\bibitem{Basso:2013vsa}
B.~Basso, A.~Sever, and P.~Vieira, ``{Spacetime and Flux Tube S-Matrices at
  Finite Coupling for N=4 Supersymmetric Yang-Mills Theory},''
  \href{http://dx.doi.org/10.1103/PhysRevLett.111.091602}{{\em Phys. Rev.
  Lett.} {\bfseries 111} no.~9, (2013) 091602},
\href{http://arxiv.org/abs/1303.1396}{{\ttfamily arXiv:1303.1396 [hep-th]}}.

\bibitem{Hodges:2009hk}
A.~Hodges, ``{Eliminating spurious poles from gauge-theoretic amplitudes},''
  \href{http://dx.doi.org/10.1007/JHEP05(2013)135}{{\em JHEP} {\bfseries 05}
  (2013) 135},
\href{http://arxiv.org/abs/0905.1473}{{\ttfamily arXiv:0905.1473 [hep-th]}}.

\bibitem{brown2009multiple}
F.~Brown, ``Multiple zeta values and periods of moduli spaces
  $\overline{\mathfrak {m}}_{0, n}$,'' in {\em Annales scientifiques de
  l'{\'E}cole Normale Sup{\'e}rieure}, vol.~42, pp.~371--489.
\newblock 2009.

\bibitem{Arkani-Hamed:2017mur}
N.~Arkani-Hamed, Y.~Bai, S.~He, and G.~Yan, ``{Scattering Forms and the
  Positive Geometry of Kinematics, Color and the Worldsheet},''
  \href{http://dx.doi.org/10.1007/JHEP05(2018)096}{{\em JHEP} {\bfseries 05}
  (2018) 096}, \href{http://arxiv.org/abs/1711.09102}{{\ttfamily
  arXiv:1711.09102 [hep-th]}}.

\bibitem{Arkani-Hamed:2019plo}
N.~Arkani-Hamed, S.~He, T.~Lam, and H.~Thomas, ``{Binary Geometries,
  Generalized Particles and Strings, and Cluster Algebras},''
  \href{http://arxiv.org/abs/1912.11764}{{\ttfamily arXiv:1912.11764
  [hep-th]}}.

\bibitem{Arkani-Hamed:2017vfh}
N.~Arkani-Hamed, H.~Thomas, and J.~Trnka, ``{Unwinding the Amplituhedron in
  Binary},'' \href{http://dx.doi.org/10.1007/JHEP01(2018)016}{{\em JHEP}
  {\bfseries 01} (2018) 016}, \href{http://arxiv.org/abs/1704.05069}{{\ttfamily
  arXiv:1704.05069 [hep-th]}}.

\bibitem{He:2018okq}
S.~He and C.~Zhang, ``{Notes on Scattering Amplitudes as Differential Forms},''
  \href{http://dx.doi.org/10.1007/JHEP10(2018)054}{{\em JHEP} {\bfseries 10}
  (2018) 054}, \href{http://arxiv.org/abs/1807.11051}{{\ttfamily
  arXiv:1807.11051 [hep-th]}}.

\bibitem{ArkaniHamed:2010gh}
N.~Arkani-Hamed, J.~L. Bourjaily, F.~Cachazo, and J.~Trnka, ``{Local Integrals
  for Planar Scattering Amplitudes},''
  \href{http://dx.doi.org/10.1007/JHEP06(2012)125}{{\em JHEP} {\bfseries 06}
  (2012) 125},
\href{http://arxiv.org/abs/1012.6032}{{\ttfamily arXiv:1012.6032 [hep-th]}}.

\bibitem{Bourjaily:2018aeq}
J.~L. Bourjaily, A.~J. McLeod, M.~von Hippel, and M.~Wilhelm, ``{Rationalizing
  Loop Integration},'' \href{http://dx.doi.org/10.1007/JHEP08(2018)184}{{\em
  JHEP} {\bfseries 08} (2018) 184},
\href{http://arxiv.org/abs/1805.10281}{{\ttfamily arXiv:1805.10281 [hep-th]}}.

\bibitem{Brown:2013qva}
F.~Brown, \href{http://dx.doi.org/10.1017/CBO9781139208642.006}{``{Iterated
  integrals in quantum field theory},''} in {\em {6th Summer School on
  Geometric and Topological Methods for Quantum Field Theory}}, pp.~188--240.
\newblock 2013.

\bibitem{CaronHuot:2012ab}
S.~Caron-Huot and K.~J. Larsen, ``{Uniqueness of two-loop master contours},''
  \href{http://dx.doi.org/10.1007/JHEP10(2012)026}{{\em JHEP} {\bfseries 10}
  (2012) 026}, \href{http://arxiv.org/abs/1205.0801}{{\ttfamily arXiv:1205.0801
  [hep-ph]}}.

\bibitem{Bourjaily:2017bsb}
J.~L. Bourjaily, A.~J. McLeod, M.~Spradlin, M.~von Hippel, and M.~Wilhelm,
  ``{Elliptic Double-Box Integrals: Massless Scattering Amplitudes beyond
  Polylogarithms},''
  \href{http://dx.doi.org/10.1103/PhysRevLett.120.121603}{{\em Phys. Rev.
  Lett.} {\bfseries 120} no.~12, (2018) 121603},
\href{http://arxiv.org/abs/1712.02785}{{\ttfamily arXiv:1712.02785 [hep-th]}}.

\bibitem{Bourjaily:2018ycu}
J.~L. Bourjaily, Y.-H. He, A.~J. McLeod, M.~von Hippel, and M.~Wilhelm,
  ``{Traintracks through Calabi-Yau Manifolds: Scattering Amplitudes beyond
  Elliptic Polylogarithms},''
  \href{http://dx.doi.org/10.1103/PhysRevLett.121.071603}{{\em Phys. Rev.
  Lett.} {\bfseries 121} no.~7, (2018) 071603},
\href{http://arxiv.org/abs/1805.09326}{{\ttfamily arXiv:1805.09326 [hep-th]}}.

\end{thebibliography}\endgroup
\end{document}